\documentclass[prd,twocolumn,amsmath,amssymb,floatfix]{revtex4}

\usepackage{bm}
\usepackage{amsmath}
\usepackage{epsf}
\usepackage{color}

\newcommand{\beq}{\begin{equation}}

\newcommand{\eeq}{\end{equation}}
\newcommand{\beqa}{\begin{eqnarray}}
\newcommand{\eeqa}{\end{eqnarray}}

\newcommand{\bk}{{\bf k}} 
\newcommand{\bx}{{\bf x}}

\newcommand{\Mobs}{M^{\rm obs}}
\newcommand{\Mtrue}{M}
\newcommand{\Mbias}{M^{\rm bias}}
\newcommand{\siglnM}{\sigma_{\ln \Mtrue}}
\newcommand{\zphot}{z^{\rm phot}}
\newcommand{\ztrue}{z}
\newcommand{\zbias}{z^{\rm bias}}
\newcommand{\sigz}{\sigma_{\ztrue}}

\newcommand{\kpar}{k_{\parallel}}
\newcommand{\kperp}{k_{\perp}}

\begin{document}

\title{Photometric Redshift Requirements for Self-Calibration of Cluster Dark Energy Studies}
\author{Marcos Lima$^{1,2}$ and Wayne Hu$^{2,3}$}
\email{mvlima@uchicago.edu}
\affiliation{
{}$^{1}$Department of Physics, University of Chicago, Chicago IL 60637\\
{}$^{2}$Kavli Institute for Cosmological Physics, University of Chicago, Chicago IL 60637\\
{}$^{3}$Department of Astronomy \& Astrophysics
and Enrico Fermi Institute, University of Chicago, Chicago IL 60637
}
\date{\today}

\begin{abstract}
\baselineskip 11pt
The ability to constrain dark energy from the evolution of 
galaxy cluster counts is limited by the imperfect 
knowledge of cluster redshifts. 
Ongoing and upcoming surveys will mostly rely on redshifts
estimated from broad-band photometry (photo-z's).
For a Gaussian distribution for the cluster photo-z errors and
a high cluster yield cosmology defined by the WMAP 1 year results,
 the photo-z bias and scatter 
needs to be known better than 0.003 and 0.03, respectively, 
in order not to degrade dark energy constrains 
by more than $10\%$ for a survey with specifications similar
to the South Pole Telescope.
Smaller surveys and cosmologies with lower cluster yields
produce weaker photo-z requirements, though relative to worse 
baseline constraints. 
Comparable photo-z requirements 
are necessary in order to employ self-calibration techniques 
when solving for dark energy and observable-mass parameters 
simultaneously.  
On the other hand, self-calibration in combination with external mass
inferences helps
reduce photo-z requirements and provides important consistency checks for 
future cluster surveys.  
In our fiducial model, training sets with spectroscopic
redshifts
for $\sim 5\%-15\%$ of the detected clusters are required
in order to keep degradations in the dark energy equation of state
lower than $20\%$.
\end{abstract}

\maketitle

\section{Introduction}

The abundance of clusters of galaxies as a function of their mass
and redshift is potentially a powerful cosmological probe. 
The sensitivity to the underlying cosmology comes 
from the dependence of the abundance on the comoving volume element and, 
more importantly, from the exponential sensitivity of the cluster 
mass function to the amplitude of 
linear density perturbations. 
Both of these features depend upon
the matter content of the universe and the underlying theory of
gravity. 
Possible applications include constraints on theories 
of modified gravity \cite{Tan06}, neutrino masses \cite{WanHaiHuKhoMay05}, 
the total matter density $\Omega_M$
and the amplitude of linear fluctuations $\sigma_8$ 
\cite{Glaetal07,Rozetal07}. 
Furthermore, cluster counts 
as a function of redshift offer a  promising technique 
to explore and constrain dark energy parameters because of the 
suppression in the growth of perturbations during the acceleration epoch. 
In practice these studies are done by comparing observations to
theoretical predictions from simulations as a function of
cosmology.

However, there are many observational challenges to the
use of clusters to constrain cosmology. 
First, there are different techniques of cluster detection,
each one attempting to obtain cluster samples as complete
and clean as possible.
Typical cluster finding methods explore signals such as the
Sunyaev-Zel'dovich (SZ) flux decrement, X-ray 
temperature, X-ray surface brightness, overdensities
in space and color from optical observations 
and the weak lensing shear.
When comparing the properties of the observed samples to
simulation predictions,
knowledge of the selection function is essential since
observational effects particular to each cluster finder
need to be properly accounted for.

Next, cluster masses must be estimated in ways that 
may or may not be tied  to the cluster finder employed. 
In general, the mass is not a direct observable and needs to be 
obtained 
through the relation between an observable
proxy and mass, the observable-mass relation.
Uncertainties in mass conversion 
can lead to degenerate effects that destroy
most of the
information in cluster counts if
not well calibrated. 

In order to overcome this degeneracy a set of so-called 
self-calibration techniques has
been developed recently.
By requiring consistency between number counts and other cluster properties,
it is possible to solve simultaneously for cosmology
and observable-mass parameters. 
This can be accomplished by follow-up of a small cluster 
sample \cite {MajMoh03}, by using information on the clustering properties
of clusters from their power spectrum \cite{MajMoh03} or their sample
covariance from counts in cells \cite{LimHu04,LimHu05}, and by using 
information from
the shape of the observed mass function \cite{Hu03a,LimHu05}.
Another approach 
uses physical models of cluster structure \cite {You06} which is similar 
in spirit to imposing
priors on observable-mass parameters.
Combining these approaches allows for tests of the assumptions 
underlying the individual approaches.  

Finally, cluster redshifts must be estimated. 
Whereas spectroscopic redshifts are very accurate,
when dealing with large data sets it becomes impractical
to obtain spectra for large fractions of objects. 
Alternatively redshifts can be estimated from 
broad-band photometry in a finite number of filter band-passes
(see e.g. \cite{Cunetal07} and 
references therein).
Photometry can be viewed as a coarse spectroscopy that
probes the most prominent spectral features as they move
from their rest positions when the object of interest is redshifted.   
Redshifts thus estimated, known as photometric
redshifts (photo-z's), can be efficiently calculated for millions
of objects and can use not only colors but any observable
that correlates with redshift \cite{OyaLimCunLinFriShe07}. 

Since the interpretation of the counts depends sensitively on an accurate
determination of the redshift distribution of the clusters and not
on the redshift precision for an individual cluster, 
photo-z's are well suited to such studies. 
Clusters detected by the SZ effect, for instance,
will typically be followed up optically so that
photo-z's can be calculated. 
Optical cluster finding algorithms can choose to derive 
photo-z's during the cluster finding process itself 
\cite{GlaYee00, Koeetal07a} or 
 use externally derived photo-z's \cite{Treetal06, Donetal07}
 with different implications for the propagation of photo-z errors.
 
In this paper,  we study how the knowledge of the cluster 
photo-z error distribution can affect the ability to 
use cluster counts to constrain dark energy.
Previous works have addressed this question for 
different cosmological probes, including 
cluster counts \cite{HutKimKraBro04}, supernova \cite{HutKimKraBro04}, 
baryon acoustic oscillations \cite{ZhaKno06,Zha06} and
weak lensing tomography \cite{HutTakBerJai06,MaHuHut06}.
In particular, Huterer et al. \cite{HutKimKraBro04} studied 
the effect of systematic shifts in centroids of redshift bins
on cosmological constraints, in the
context of perfect knowledge of cluster masses.
Here we generalize that analysis by considering 
the full redshift error distribution and allowing 
redshift bias and scatter parameters to be 
arbitrary functions of redshift. 
We also consider the photo-z requirements necessary for 
self-calibration of the observable-mass relation
when one is simultaneously solving  for cluster masses 
and cosmology.

We start in $\S$\ref{sec_cou} and $\S$\ref{sec_cov} describing 
how redshift errors affect cluster number
counts and their sample covariance respectively. 
In $\S$\ref{sec_fis} we describe
the fiducial models assumed and the Fisher matrix formalism,
which is employed in $\S$\ref{sec_res} to study how
redshift errors degrade dark energy constraints in 
various cases of interest. 
Finally, in $\S$\ref{sec_dis} we discuss the results and conclude.

\section{Number Counts} \label{sec_cou}

For a given cosmology, simulations predict the
comoving number density of dark matter halos as 
a function of mass and redshift.
For illustrative purposes, we will identify these halos with 
clusters and employ a fit to simulations for the halo 
differential comoving number density \cite{Jenetal01}
\begin{equation}
{d \bar n \over d\ln M} = 0.3 {\rho_{m} \over M} {d \ln \sigma^{-1} \over d\ln M}
        \exp[-|\ln \sigma^{-1} + 0.64|^{3.82}]\,,
\label{eqn:massfun}
\end{equation}
\noindent where $\sigma^2(M;z)\equiv \sigma^2_{R}(z)$, the linear density 
field variance in a region enclosing $M=4\pi R^3\rho_m/3$ 
at the mean matter density today $\rho_{m}$. 
Because shifts in cluster masses can modify cluster counts mimicking 
a change in cosmology, the exponential sensitivity to $\sigma$
 will only be a benefit in 
practice if the observable-mass distribution is also well known. 
Fortunately observations and simulations suggest observable-mass scaling 
relations that can be parametrized in simple forms and allow for 
reasonable degree of calibration \cite{KraVikNag06,Nag06,Ohaetal06}.
Following \cite{LimHu05}, we take the probability of assigning a 
mass $\Mobs$ to a cluster of true mass $M$ to be given by a Gaussian 
distribution in $\ln M$
\begin{equation}
p(\Mobs | \Mtrue) = {1 \over \sqrt{2\pi \siglnM^2} } \exp\left[ -x^2(\Mobs) \right] \,,
\end{equation}
\noindent where
\begin{equation}
x(\Mobs) \equiv { \ln \Mobs - \ln M - \ln  \Mbias \over \sqrt{2 \siglnM^2}} \,.
\end{equation}

For simplicity, we will allow the mass bias $\ln \Mbias$ 
and the variance $\siglnM^2$ to vary with redshift but not mass.
The redshift dependent average number density of clusters within the 
observable mass range $\Mobs_{\alpha} \le \Mobs \le \Mobs_{\alpha+1}$ 
is given by
\begin{align}
\bar n_{\alpha}(z) & \equiv \int_{\Mobs_{\alpha}}^{\Mobs_{\alpha+1}} {d \Mobs \over \Mobs}
\int {dM \over M} { d \bar n \over d\ln M}
p(\Mobs | \Mtrue)\nonumber \\
& = \int {d M \over M} { d \bar n \over d\ln M} {1\over 2} \left[ {\rm erfc}(x_{\alpha}) - {\rm erfc}(x_{\alpha+1}) \right] \,,
\label{eqn:binneddensity}
\end{align}
\noindent where $x_{\alpha} = x(\Mobs_{\alpha})$.  
Notice that this corresponds to the cumulative number density above 
some sharp mass threshold in the limit that 
$\siglnM^{2} \rightarrow 0$ and $\Mobs_{\alpha+1} \rightarrow \infty$.
The mean number of clusters is then obtained by integrating
the mean number density in the redshift dependent comoving volume 
element $d^3x$.
Let us use spherical coordinates to parametrize 
the spatial position vector as ${\bf x}=(r,\theta,\phi)$, 
where $r(z)$ is the 
angular diameter distance to redshift $z$ and 
$(\theta,\phi)$ 
parametrize the solid angle $\Omega$ such that
$d \Omega = \sin \theta d \theta d\phi$. 
Since we will only consider flat cosmologies, 
$r(z)$ coincides with the comoving distance. The
volume element is given by  
\begin{eqnarray}
d^3x =  
r^2drd\Omega
=\frac{r^2(z)}{H(z)}dz d\Omega \,,
\label{eqn:vol}
\end{eqnarray}
\noindent where 
$H(z)$ is the Hubble parameter to redshift $z$. 
Redshift uncertainties affect the redshift bin size as well as the 
observed angle, distorting the volume element and changing the number counts. 
We take the probability of assigning a photo-z $\zphot$ to a cluster  
of true redshift $z$ to be also a Gaussian distribution
\begin{eqnarray}
p(\zphot| z) &=& {1 \over \sqrt{2\pi \sigz^2  } } \exp\left[ -y^2(\zphot) \right] \,,
\label{eqn:zdist_gauss}
\end{eqnarray}
\noindent where
\begin{eqnarray}
y(\zphot)&\equiv& { \zphot    - z - \zbias      \over \sqrt{2 \sigz^2  }} 
\end{eqnarray}
\noindent and the redshift bias $\zbias=\zbias(z)$ and variance 
$\sigz^2=\sigz^2(z)$ are allowed to be arbitrary functions of redshift.

\begin{figure*}[tb]
  \begin{minipage}[t]{6.8in}
    \centerline{\epsfxsize=6.8in\epsffile{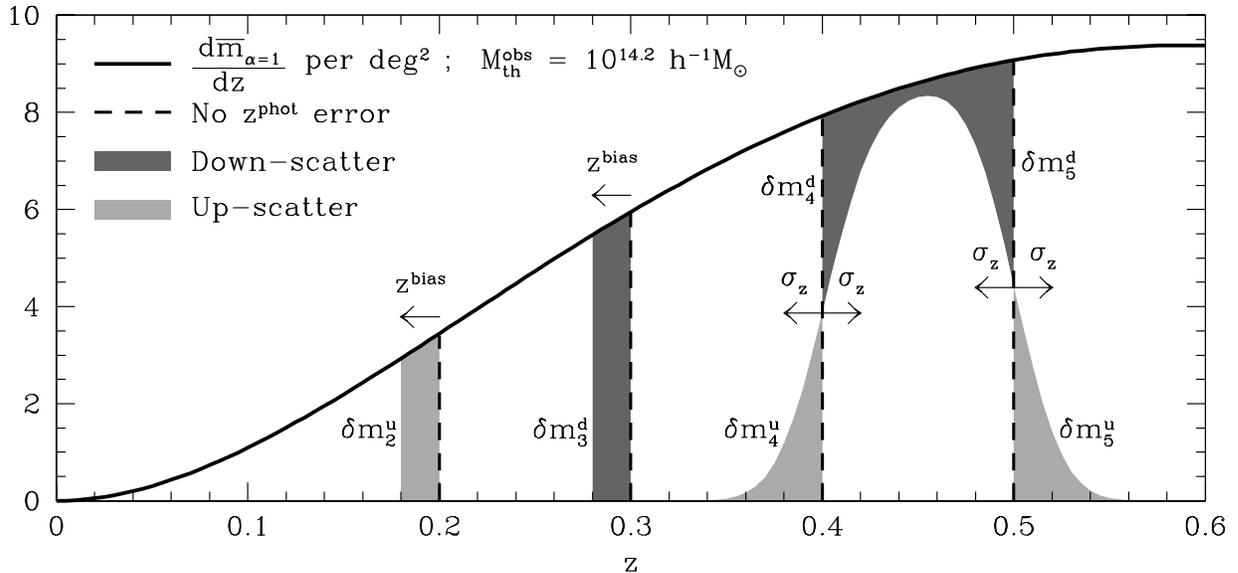}}
  \end{minipage}
\caption{\footnotesize
The redshift distribution of cluster counts per deg$^2$ above a 
sharp mass threshold
$M^{\rm obs}_{\rm th}=10^{14.2}h^{-1}M_{\odot}$ is shown (solid line)
for the fiducial survey properties defined in \S \ref{sec_fis} in
the WMAP1 cosmology. 
The effects of photo-z bias and scatter are shown for 
$\zbias=\sigz=0.02$ at redshift bins $[0.2,0.3]$ and $[0.4,0.5]$ 
respectively.
The dashed lines indicate a perfect redshift
selection in the absence of photo-z uncertainties. 
Photo-z errors scatter objects up (light gray) and down 
(dark gray), changing the observed number counts. 
In the limit that $\sigz^2\rightarrow 0$, a positive 
redshift bias decreases counts in regions where $dm/dz$ 
is increasing and vice-versa; the sensitivity of the counts is then 
controlled by $-d^2m/dz^2$ at the bin.
The effect of scatter in each bin border scales 
as $\sim 0.5d^2m/dz^2$ in this limit;
the total effect then depends on
how $d^2m/dz^2$ varies from bin to bin and is
proportional to $0.5d^3m/dz^3$.}
\label{fig:dmdz}
\end{figure*}
Given a perfect angular selection characterized by an angular top
hat window $W^{\rm th}_{i}(\Omega)$, the mean number of clusters 
in a photo-z bin defined by $\zphot_i \le \zphot \le \zphot_{i+1}$ 
is
\begin{eqnarray}
\bar m_{\alpha,i} &=& \int_{\zphot_i}^{\zphot_{i+1}} d\zphot 
                  \int d^3x \bar n_{\alpha} W_{i}^{\rm th}(\Omega) p(\zphot| z) \nonumber \\ 
                  &=&\int d^3x \bar n_{\alpha} W_i({\bf x}) \,,
\label{eqn:numwindow}
\end{eqnarray}
\noindent where the total window function $W_i({\bf x})$ is given by 
\begin{eqnarray}
W_i({\bf x})=W_i^{\rm th}(\Omega)F_i(z) \,,
\label{eqn:totalwindow}
\end{eqnarray}
\noindent and
\begin{eqnarray}
F_i(z)&=&\int_{\zphot_i}^{\zphot_{i+1}} d\zphot p(\zphot| z) \nonumber \\
      &=&{1\over 2} \left[ {\rm erfc}(y_i) - {\rm erfc}(y_{i+1}) \right] \,,
\label{eqn:zselection}
\end{eqnarray}
\noindent with $y_i=y(\zphot_i)$. 
The total window function in Eq.~(\ref{eqn:totalwindow}) takes into 
account photo-z error information carried by $F_i(z)$. 
Notice that when $\zbias=\sigz=0$, the distribution
$p(\zphot|z)=\delta(\zphot-z)$ and $F_i(z)$ becomes a top hat 
$W_i^{\rm th}(z)$, 
which combined with the angular top hat produces the full 
$3$d top hat window 
$W_i^{\rm th}({\bf x})=W_i^{\rm th}(\Omega)W_i^{\rm th}(z)$
of a perfect selection. 
Using Eqs.~(\ref{eqn:vol}),~(\ref{eqn:numwindow}),~(\ref{eqn:totalwindow}) 
and (\ref{eqn:zselection}) 
the mean number counts in a cell defined by
the photo-z bin and a solid angle $\Delta \Omega$ can be written as 
\begin{eqnarray} 
\bar m_{\alpha,i}&=& \int d\Omega dz \frac{r^2(z)}{H(z)} \bar n_{\alpha}(z) W_i^{\rm th}(\Omega) F_i(z)\nonumber \\
           &=& \int dz \frac{d \bar m_{\alpha}}{dz}
               {1\over 2} \left[ {\rm erfc}(y_i) - {\rm erfc}(y_{i+1}) \right] \,,
\label{eqn:binnedcounts}
\end{eqnarray}
\noindent where we used the angular top hat window to perform the solid angle
integral and defined
\begin{eqnarray}
\frac{d \bar m_{\alpha}}{dz}\equiv \Delta \Omega \frac{r^2(z) \bar n_{\alpha}(z)}{H(z)} \,.
\label{eqn:zdistdef}
\end{eqnarray}
The solid angle $\Delta \Omega$ is related to the angular extension $\theta_s$ 
of the cell by
\begin{eqnarray}
\Delta \Omega  = \int d\Omega W_i^{\rm th}(\Omega) 
                = 2\pi(1-\cos \theta_s)
              \approx \pi \theta_s^2
\end{eqnarray}
\noindent and the approximation is true in the flat sky regime, valid
for small angle windows.

It is instructive to consider the  sensitivity of number counts
to the photo-z parameters in some limits. 
Let us consider the simple case of a sharp mass threshold 
(i.e. drop the $\alpha=1$ index) and let us assume that the 
redshift bias and variance are smooth functions of redshift. 
From Eq.~(\ref{eqn:binnedcounts}), the relative sensitivity of 
the mean counts $\bar m_i= \bar m_{\alpha=1,i}$ in a photo-z 
bin of width $\delta z^p_i=\zphot_{i+1}-\zphot_{i}$ to the 
redshift bias around $\sigz^2=0$ is
\begin{eqnarray}
\lim_{\sigz^2\rightarrow 0} \frac{\partial \ln \bar m_{i}} {\partial \zbias} = 
\frac{1}{\bar m_{i}}\frac{d \bar m}{dz} \Big|_{y_{i+1}=0}^{y_{i}=0} \sim 
-\frac{1}{\bar m_{i}}\frac{d^2 \bar m}{dz^2}\delta z^p_i\,,
\label{eqn:zbiassens}
\end{eqnarray}
\noindent whereas the sensitivity to the redshift variance is
\begin{eqnarray}
\lim_{\sigz^2\rightarrow 0} \frac{\partial \ln \bar m_{i}} {\partial \sigz^2} = 
-\frac{1}{2 \bar m_{i}}\frac{d^2 \bar m}{dz^2} \Big|_{y_{i+1}=0}^{y_{i}=0} \sim 
\frac{1}{2 \bar m_{i}}\frac{d^3 \bar m}{dz^3}\delta z^p_i\,,
\label{eqn:sigz2sens}
\end{eqnarray}
\noindent and the derivatives are evaluated at the photo-z bin center 
$z^p_i=(\zphot_i+\zphot_{i+1})/2$. 
Note that these derivatives are insensitive to the actual value of 
$\sigma_z^2$  in this limit.   Consequently we will employ 
the variance $\sigma_z^2$ instead of the rms $\sigma_z$ in 
the forecasts below.

In general, the sensitivity of the counts to higher moments of the 
photo-z error distribution will depend on higher derivatives
of the true redshift distribution and will tend to be less important. 
Fig.~\ref{fig:dmdz} shows the redshift distribution given by 
Eq.~(\ref{eqn:zdistdef}) plotted as a function of redshift for the 
fiducial model defined in $\S$ \ref{sec_fis} as well as
the effects of $\zbias$ and $\sigz$ on 
the observed number counts.
The distribution is shown only up to $z=0.6$ for clarity.

The result of Eq.~(\ref{eqn:zbiassens}) is intuitive from the fact 
that a positive redshift bias decreases the counts in regions 
where the redshift distribution is increasing and vice-versa. 
Consider the effect of the redshift bias on the number counts in the 
redshift bin $[z_2,z_3]=[0.2,0.3]$ as shown in Fig.~\ref{fig:dmdz}. 
We have that 
$\partial \bar m_2/\partial \zbias \sim (\delta m^u_2-\delta m^d_3)/\zbias$,
where $\delta m^u_2$ is number of upscattered clusters around $z_2$ 
and $\delta m^d_3$ is the number of downscattered clusters around $z_3$. 
Approximating these by rectangles, we can combine
\begin{eqnarray}
\delta m^u_2 \sim \frac{d \bar m}{dz}\Big|^{z_2}\zbias \, ,
\end{eqnarray} 
\noindent with a similar expression for $\delta m^d_3$, and obtain 
Eq.~(\ref{eqn:zbiassens}). 

The sensitivity to the redshift variance allows a similar 
interpretation. In Fig.~\ref{fig:dmdz} we consider the effect of 
the redshift scatter on the counts in the 
redshift bin $[z_4,z_5]=[0.4,0.5]$. In this case we have
$\partial \bar m_4/\partial \sigz^2 
\sim (\delta m^u_4-\delta m^d_4+\delta m^u_5-\delta m^d_5)/\sigz^2$. 
Approximating these now by triangles, we have
\begin{eqnarray}
\delta m^u_4 \sim \frac{1}{2}\left[0.5\frac{d \bar m}{dz}\Big|^{z_4-\sigz}\right]\sigz \, .
\end{eqnarray}
\noindent A similar expression for $\delta m^d_4$ at $z_4+\sigz$ gives
\begin{eqnarray} 
\frac{(\delta m^u_4-\delta  m^d_4)}{\sigz^2} 
\sim \frac{0.5}{2\sigz} \frac{d \bar m}{dz} \Big|_{z_4+\sigz}^{z_4-\sigz} 
\sim -0.5\frac{d^2 \bar m}{dz^2}\Big|^{z_4}.
\end{eqnarray}

The corresponding result for $(\delta m^u_{5}-\delta m^d_{5})/\sigz^2$ then
leads to Eq.~(\ref{eqn:sigz2sens}).

The limits in Eqs.~(\ref{eqn:zbiassens}) and 
(\ref{eqn:sigz2sens}) are useful to get qualitative insight into the
effect of redshift bias and variance when they are small and
vary smoothly in redshift.
Quantitatively, the finite values of these parameters 
as well as more general parametrization of these functions
may cause the sensitivity of the counts to change from these special limits.
In \S \ref{sec_res}, we explore how the sensitivity of the counts to photo-z 
parameters ultimately affect dark energy constraints.

\section{Sample Covariance} \label{sec_cov}

The number counts $m_{\alpha,i}({\bf x})$ fluctuate in space tracing 
the linear density fluctuations $\delta({\bf x})$ induced by large
scale structure
\begin{eqnarray}
m_{\alpha,i}({\bf x}) &=& \bar m_{\alpha,i}[ 1 + b_{\alpha}(z) \delta({\bf x}) ] \,,
\label{eqn:count_fluc}
\end{eqnarray}
\noindent where $b_{\alpha}(z)$ is the 
average cluster linear bias predicted from 
the distribution in Eq.(\ref{eqn:binneddensity}) 
\begin{eqnarray}
b_{\alpha}(z) = {1 \over \bar n_{\alpha}(z)}  \int {d M\over M}  {d \bar n_{\alpha}(z) \over d\ln M} b(M;z)\,,
\end{eqnarray}
\noindent and we take a fit to simulations of \cite{SheTor99}
\begin{equation}
b(M;z) = 1 + {a_c \delta_c^2/\sigma^2 -1 \over \delta_c} 
         + { 2 p_c \over \delta_c [ 1 + (a \delta_c^2/\sigma^2)^{p_c}]}
\label{eqn:bias}
\end{equation}
\noindent with $a_c=0.75$,  $p_c= 0.3$,  and $\delta_c=1.69$.
From Eq.~(\ref{eqn:count_fluc}), the counts $m_{\alpha,i}$ then possess 
a sample covariance given by \cite{HuKra02}
\begin{eqnarray}
S^{\alpha \beta}_{ij} &=&\langle (m_{\alpha,i} -\bar m_{\alpha,i})(m_{\beta,j} - \bar m_{\beta,j})\rangle \label{eqn:covariance} \\
&=& b_{\alpha,i} \bar m_{\alpha,i} b_{\beta,j} \bar m_{\beta,j} \int{d^3 k \over (2\pi)^3} W_i^*(\bk)W_j(\bk) P(k)\,,\nonumber 
\end{eqnarray}
\noindent which accounts for the clustering of clusters due to large
scale structure. Here $ W_i(\bk)$ is the Fourier transform of the window function
and for simplicity we assumed that $b_{\alpha}(z)$ does not vary 
considerably within the photo-z bin $i$ and can be approximated by 
$b_{\alpha}(z) \approx b_{\alpha}(z^p_i) \equiv  b_{\alpha,i}$.
In the more general case where $b_{\alpha}(z)$
varies considerably within the photo-z bin, $W_i(\bk)$ above would
be the Fourier transform of $W_i(\bx)b_{\alpha}(z)$, and it would be
harder to find an analytical expression for it. 
We chose not to consider this
case since the approximation holds well for the small photo-z 
bin size ($\delta z^p =0.1$) considered in our fiducial model.
  
Let the photo-z bin $i$ be at angular diameter distance $r_i \equiv r(z^p_i)$
and have width $\delta r_i \equiv r(\zphot_{i+1})-r(\zphot_{i})$.
Under the assumption
 that $H(z)$, $\zbias(z)$ and $\sigz(z)$ also do not change appreciably 
inside the bin, so that 
$H(z) \approx H(z^p_i) \equiv  H_{i}$ and likewise for 
$\zbias(z^p_i) \equiv \zbias_i$ and $\sigz(z^p_i) \equiv \sigma_{z,i}$,
the window 
$W_i(\bk)$ is given by (see Appendix \ref{app_win})
\begin{eqnarray}
W_i(\bk)=&&2\exp{\left[ i \kpar \left( r_i+\frac{\zbias_i}{H_i} \right) \right] }
              \exp{ \left[- \frac{\sigma_{z,i}^2 \kpar^2}{2H_i^2}  \right] } \nonumber \\
           &&   \frac{\sin( \kpar \delta r_i/2) }{\kpar \delta r_i/2} 
              \frac{J_1(\kperp r_i \theta_s)}{\kperp r_i \theta_s}\,. 
\label{eqn:window}
\end{eqnarray}
\noindent Since $\zbias_i \ll \delta z^p_i$ in all practical cases, 
$b_{\alpha,i}$ and $P(k)$ do not carry any strong dependence on 
the photo-z bias. 
In this case, 
Eqs.~(\ref{eqn:covariance}) and (\ref{eqn:window}) indicate that
the sample variance $S_{ii}$ does not depend on $\zbias_i$ and is
exponentially sensitive to $\sigma_{z,i}^2$.  
This will bring interesting effects when using sample covariance as a 
signal for self-calibration (see \S \ref{sec_res}).
From Eq.~(\ref{eqn:window}), we see that in the absence of photo-z errors, 
the window
function would suppress modes along the line of sight with wavelengths 
$\lambda_{\parallel} < \delta r=\delta z^p/H$. 
The presence of photo-z scatter further suppresses modes 
$\lambda_{\parallel} < \pi \sigz/H$. 
For our fiducial redshift binning of $\delta z^p=0.1$ we 
expect significant effects to appear then if 
$\sigz > \delta z^p/\pi \sim 0.03$. 
As we will see in \S \ref{sec_res}, this is roughly the value of scatter 
uncertainty where dark energy degradations start to increase 
considerably.

\section{Fisher Matrix and Self-Calibration} \label{sec_fis}

Given a parametrized model that predicts the number counts
and their sample covariance, along with a fiducial choice
for the true values of these parameters, the Fisher matrix formalism allows
us to study the impact of redshift uncertainties on dark energy constraints.
 
As discussed below, we will consider counts not only in photo-z bins, but
also in angular cells and observed mass bins. 
To simplify the notation, from now on $i$ will index a generalized pixel
of redshift, angle and mass whenever these are appropriate and $S_{ij}$ 
will be the corresponding sample covariance.
The total covariance is the sample covariance plus shot variance
\begin{eqnarray}
C_{ij} = S_{ij} + \bar m_i \delta_{ij} \,.
\end{eqnarray}

For convenience, we arrange the counts per pixel $i$ into a vector 
${\bf m} \equiv (m_1,\ldots,m_{N_{\rm pix}})$ and correspondingly 
their sample and total covariances into matrices ${\bf S}$ and ${\bf C}$.
The Fisher matrix quantifies the information in the counts on a set of 
parameters $p_\alpha$ as
\cite{HolHaiMoh01,LimHu04,HuCoh06}
\begin{eqnarray}
F_{\alpha\beta}&=&  \bar{\bf m}^t_{,\alpha} {\bf C}^{-1}
 \bar{\bf m}_{,\beta} 
+ {1\over 2} {\rm Tr} [{\bf C}^{-1} {\bf S}_{,\alpha}
 {\bf C}^{-1} {\bf S}_{,\beta} ]\,.
 \label{eqn:fisher}
\end{eqnarray}
\noindent The first term represents the information from the mean counts and 
the second term carries the information from the sample covariance of the 
counts. 
When using only counts information, we will be using only the first
term in the Fisher matrix definition, while when using the sample covariance
to employ self-calibration we will use both terms.
 The Fisher matrix approximates the covariance matrix of the parameters 
$C_{\alpha\beta} \approx [{\bf F}^{-1}]_{\alpha\beta}$
such that the marginalized error on a single parameter is
 $\sigma(p_\alpha) = [{\bf F}^{-1}]_{\alpha\alpha}^{1/2}$.  
When considering prior information on parameters of a given 
$\sigma(p_\alpha)$ we add to the Fisher matrix a contribution of 
$\sigma^{-2}(p_\alpha) \delta_{\alpha\beta}$ before inversion. 

Next we define our fiducial choices for survey properties and values for
cosmological parameters as well as nuisance parameters describing the 
observable-mass relation and photo-z's.

\subsection{Fiducial Model} 

We will take a fiducial cluster survey with specifications similar to 
the South Pole Telescope (SPT) Survey ~\cite{Ruh04}: 
an area of 4000 deg$^{2}$ and a sensitivity corresponding to a 
constant mass threshold 
 $\Mobs_{\rm th} = 10^{14.2} h^{-1} M_{\odot}$. 
We divide the number counts into photo-z bins of $\delta z^p_i =0.1$ out 
to $\zphot_{\rm max}=2$. 
When using self-calibration from sample variance information, we further
divide these counts into 400 angular cells of 10 deg$^{2}$. 
Finally, when employing self-calibration from shape of the observed mass 
function, we additionally divide the counts in 5 bins of 
$\Delta \log_{10} \Mobs = 0.2$. 
In \S \ref{subsec_fid} we will consider some variations on these fiducial
choices.

The observable-mass relation will be parametrized by the mass bias $\Mbias$ 
and variance $\siglnM^2$ as a function of redshift. 
Because the evolution in cluster parameters is expected to be smooth in 
redshift, we will assume in our fiducial model that the mass bias has a 
power law evolution given by
\begin{equation}
\ln \Mbias(z) = A_b + n_b \ln (1+z) 
\label{eqn:biaspowerlaw}
\end{equation}
\noindent with fiducial values $A_b=n_b=0$. 
For the mass variance $\siglnM^{2}$ we assume a Taylor expansion around $z=0$ 
\begin{equation}
\siglnM^{2}(z) = \siglnM^{2} |_{\rm fid} + \sum_{a=0}^{N_\sigma-1} B_a z^a
\label{eqn:variancetaylor}
\end{equation}
\noindent with fiducial values $\siglnM^{2} |_{\rm fid}=(0.25)^2$ and 
$B_a=0$. 
We have checked that using a different fiducial value of
$\siglnM^{2} |_{\rm fid}=(0.05)^2$ does not change the results. 
For the fiducial model we will take a cubic ($N_\sigma=4$) evolution of the 
mass variance, so our fiducial model will be comprised of $6$ nuisance 
parameters for the observable-mass distribution. 
In \S \ref{subsec_fid} we also consider a more general parametrization 
with one value of $\Mbias$ and $\siglnM^2$ in each photo-z bin.

For the photo-z model 
we will take the values of $\zbias(z)$ and $\sigz^2(z)$ in $N_z$ 
different redshifts as our nuisance photo-z parameters.
Values of photo-z parameters at arbitrary values of $z$ are then 
obtained by cubic spline interpolation.
Current photo-z methods can reach accuracies of $\sigz \sim 0.1$ 
for field galaxies up to moderate redshifts ($z \sim 1.4$) when
using optical filter band-passes \cite{Cunetal07} (see also 
\cite{OyaLimCunLinFriShe07} for a case with lower redshifts). 
If complemented by near-infrared filters, the accuracy can be 
improved and well controlled up to high redshifts ($z \sim 2.0$).
Photo-z's of red early-type galaxies are typically even
better because of their very prominent $4000${\rm  \AA} break. 
Finally, cluster photo-z's are expected to be further improved
by averaging the photo-z's of individual cluster galaxies.
However, this will only be the case if the presence of interlopers is 
small and well understood.
We expect therefore not only the photo-z errors to be smaller
than the corresponding errors for field galaxies, but the
error distribution to be more well behaved. 
For instance the Gaussian assumption of Eq.~(\ref{eqn:zdist_gauss}) 
should be a good approximation for clusters with well understood 
selections. 
For our fiducial photo-z model, we will take $N_z=20$, i.e. one
value of $\zbias$ and $\sigz^2$ for each bin 
$\delta z=0.1$ of {\it true} redshift. 
We will assume fiducial values $\zbias(z) |_{\rm fid}=0$ and
\begin{equation}
\sigz(z) |_{\rm fid} = 0.03(1+z)
\label{eqn:sigzlinear}
\end{equation}
\noindent to account for the fact that the photo-z scatter
is expected to increase at high redshifts.
In \S \ref{subsec_fid} , we consider a different photo-z model 
with a constant $\sigz(z) |_{\rm fid} = 0.02$, and show
that the results are very similar to the ones in the fiducial model.
That the results are insensitive to the fiducial values assumed in a given
parametrization is a check of the validity of our Fisher matrix
approach.
Obviously, changes in the parametrization and in the number of nuisance 
parameters affect the results.
The choice of $N_z=20$, corresponding to one value of $\zbias$ 
and $\sigz^2$ per photo-z bin used to count clusters 
(recall $\delta z^p_i=0.1$) is, although natural, arbitrary. 
In principle, we could choose $N_z=10$ or $40$ and keep the same photo-z
binning or alternatively use coarser/finer photo-z bins. 
The photo-z bin size is ultimately dictated by the photo-z precision that 
can be achieved, e.g. the photo-z bin size should be larger than 
$\zbias$ or $\sigz$. 
The value of $\delta z^p_i=0.1$ chosen reflects conservative expectations for 
the photo-z precision of future cluster surveys. 
In \S \ref{subsec_fid} we consider alternative models with 
$N_z=10$ and $\delta z^p_i=0.2$. 

For the fiducial cosmological parameters, we will assume a flat
universe and take parameter values based on the results of the Wilkinson
Microwave Anisotropy Probe first year data release (WMAP1). 
In \S \ref{subsec_fid} we will also consider a fiducial cosmology based 
on the third year data release (WMAP3).
We chose the WMAP1 as the fiducial case since this cosmology will place
the stronger requirements on the photo-z errors. 
For the WMAP1 case, the cosmological parameters are 
the normalization of the initial curvature spectrum 
$\delta_\zeta (=5.07\times 10^{-5})$ at $k=0.05$ Mpc$^{-1}$
(corresponding to $\sigma_8=0.91$, see \cite{HuJai03}), its tilt
$n (=1)$, the baryon density
$\Omega_bh^2 (=0.024)$, the dark matter density
$\Omega_m h^2 (=0.14)$,
and the two dark energy
parameters of interest: its density
$\Omega_{\rm DE} (=0.73)$ relative to critical
and equation of state $w(=-1)$ which we  assume to be constant.    
The fiducial values are given in parentheses.
For the WMAP3 cosmology we take 
$\delta_\zeta (=4.53\times 10^{-5})$ at $k=0.05$ Mpc$^{-1}$
(corresponding to $\sigma_8=0.76$),
$n (=0.958)$,
$\Omega_bh^2 (=0.0223)$,
$\Omega_m h^2 (=0.128)$,
$\Omega_{\rm DE} (=0.76)$ and
$w(=-1)$.
We will assume $1\%$ priors on all cosmological parameters except
for the dark energy parameters, which will vary freely. 

\section{Results} \label{sec_res}

Now we apply the formalism developed in \S \ref{sec_fis} to study 
how dark energy constraints are affected by redshift uncertainties. 
We first consider the case where masses are perfectly known and only 
information from counts is used to constrain dark energy. 
We then progress to a more pessimistic case where masses are unknown 
and self-calibration is employed in a joint fit of dark energy and 
observable-mass parameters.
We then consider the case where masses are known, and 
self-calibration is used to reduce the redshift requirements.
Finally, we study how stable our results are to changes in 
our fiducial parametrization and estimate the requirements 
on the size of calibration training sets necessary in order 
not to degrade dark energy constraints appreciably.

Given the values of the prior uncertainties in the redshift bias 
$\sigma(\zbias)$ and variance $\sigma(\sigz^2)$, which produce the 
constraint 
$\sigma(w)=\sigma\left[ w| \sigma(\zbias),\sigma(\sigz^2) \right]$ 
on the dark energy equation of state $w$, we define the $w$ degradation 
$d_w$ relative to a reference case with $w$ constraint 
$\sigma(w)|_{\rm ref} < \sigma(w)$ as  
\begin{eqnarray}
d_w(\sigma(\zbias),\sigma(\sigz^2))=
\frac{ \sigma\left[ w| \sigma(\zbias),\sigma(\sigz^2) \right] }
     { \sigma(w)|_{\rm ref} } -1  
\label{eqn:degra}
\end{eqnarray}
\noindent and likewise for the $\Omega_{\rm DE}$ degradation 
$d_{\Omega_{\rm DE}}$.
Unless otherwise specified, we will take the reference case to be 
the baseline case of perfect redshifts 
$\sigma(w)|_{\rm ref}=\sigma\left[ w| 0, 0 \right]$.
In some cases though we will also use this degradation definition 
to generally compare two different scenarios such as
the WMAP1 versus WMAP3 cosmologies and implementations assuming 
masses to be perfectly known versus self-calibrated.

\subsection{Perfect Mass Calibration} \label{subsec_per.mas}

\begin{figure*}[tb]
  \begin{minipage}[t]{3.4in}
    \centerline{\epsfxsize=3.4in\epsffile{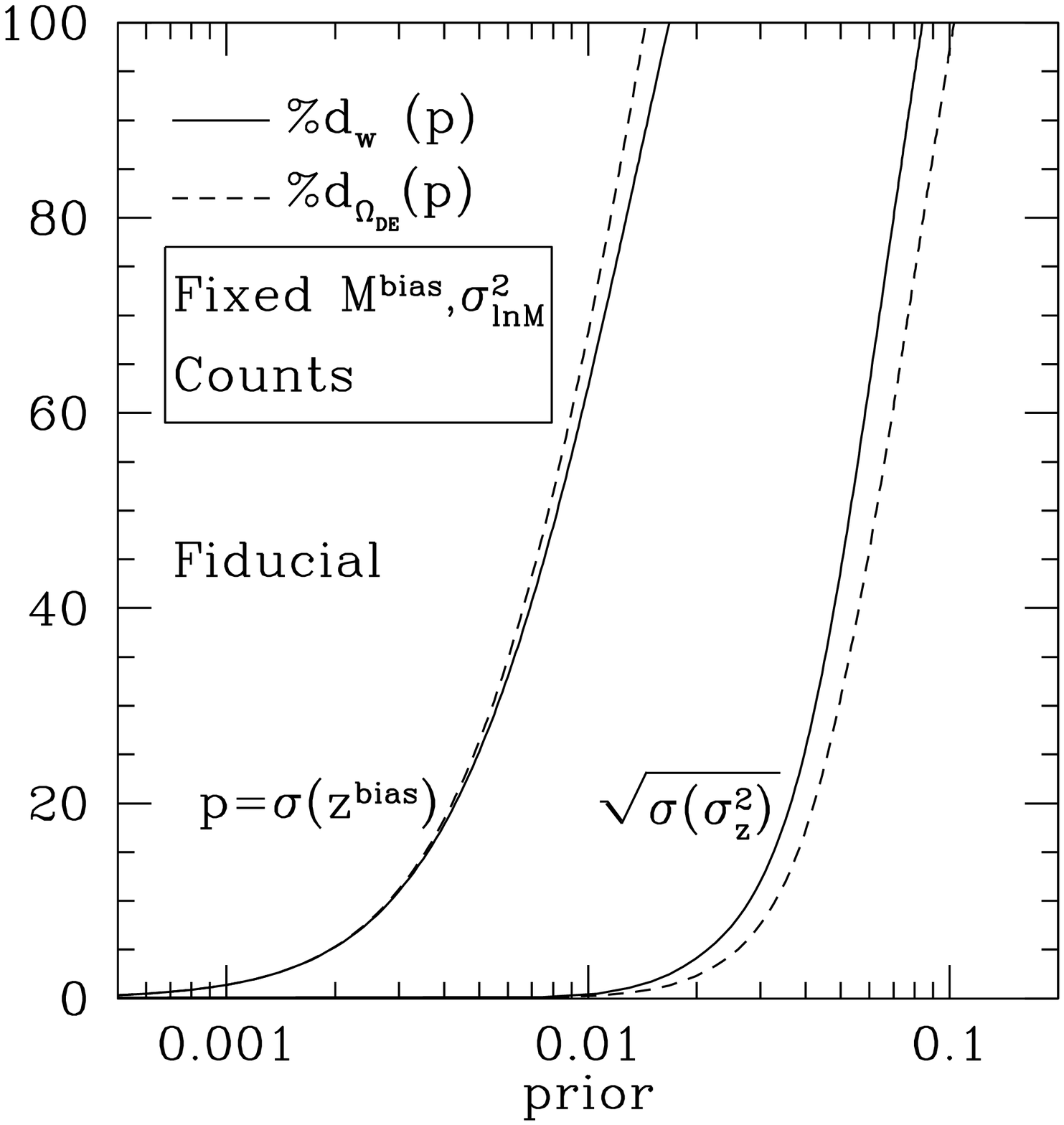}}
  \end{minipage}
  \hfill
  \begin{minipage}[t]{3.4in}
    \centerline{\epsfxsize=3.4in\epsffile{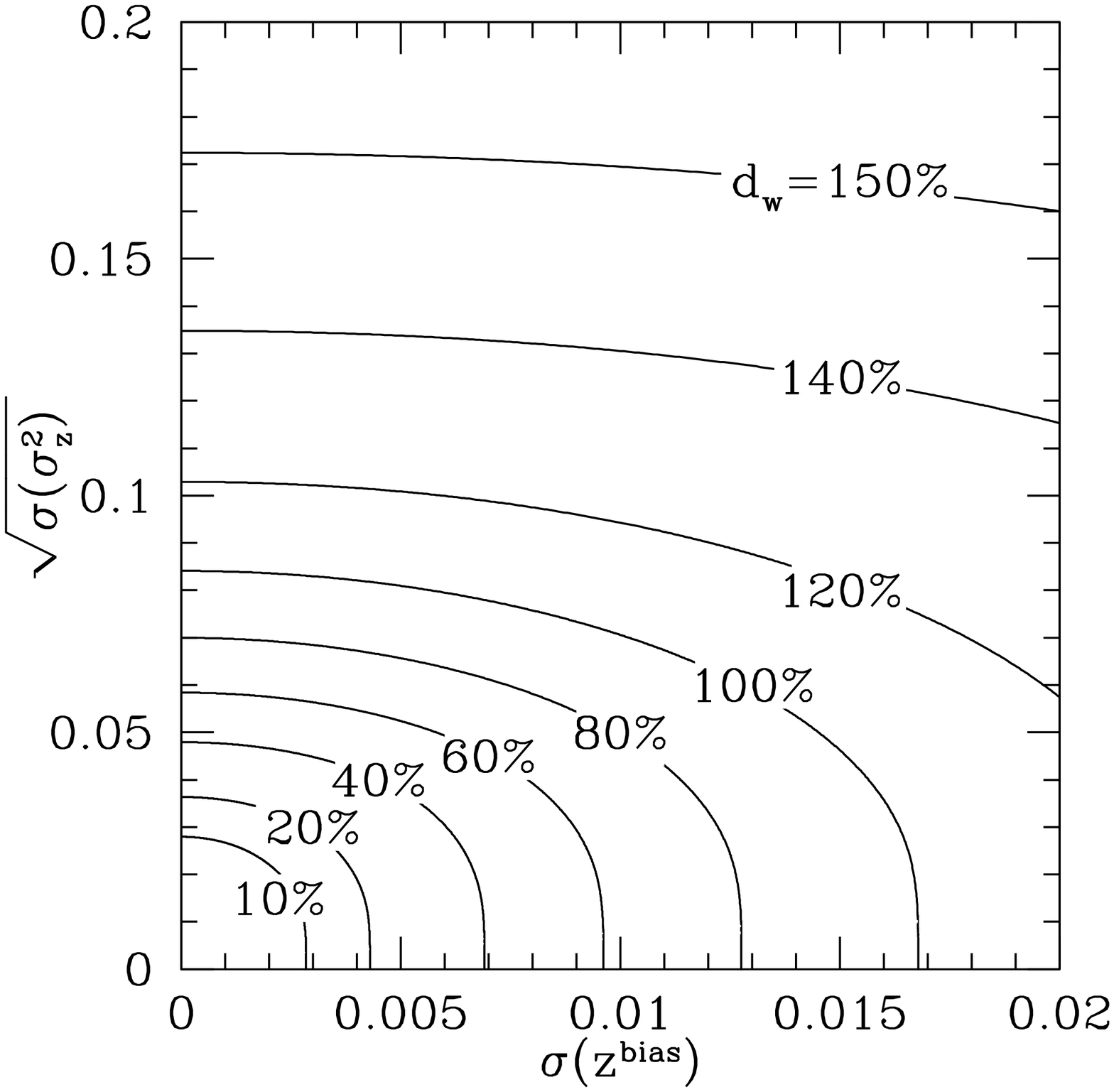}}
  \end{minipage}
  \begin{minipage}[t]{3.4in}
    \centerline{\epsfxsize=3.4in\epsffile{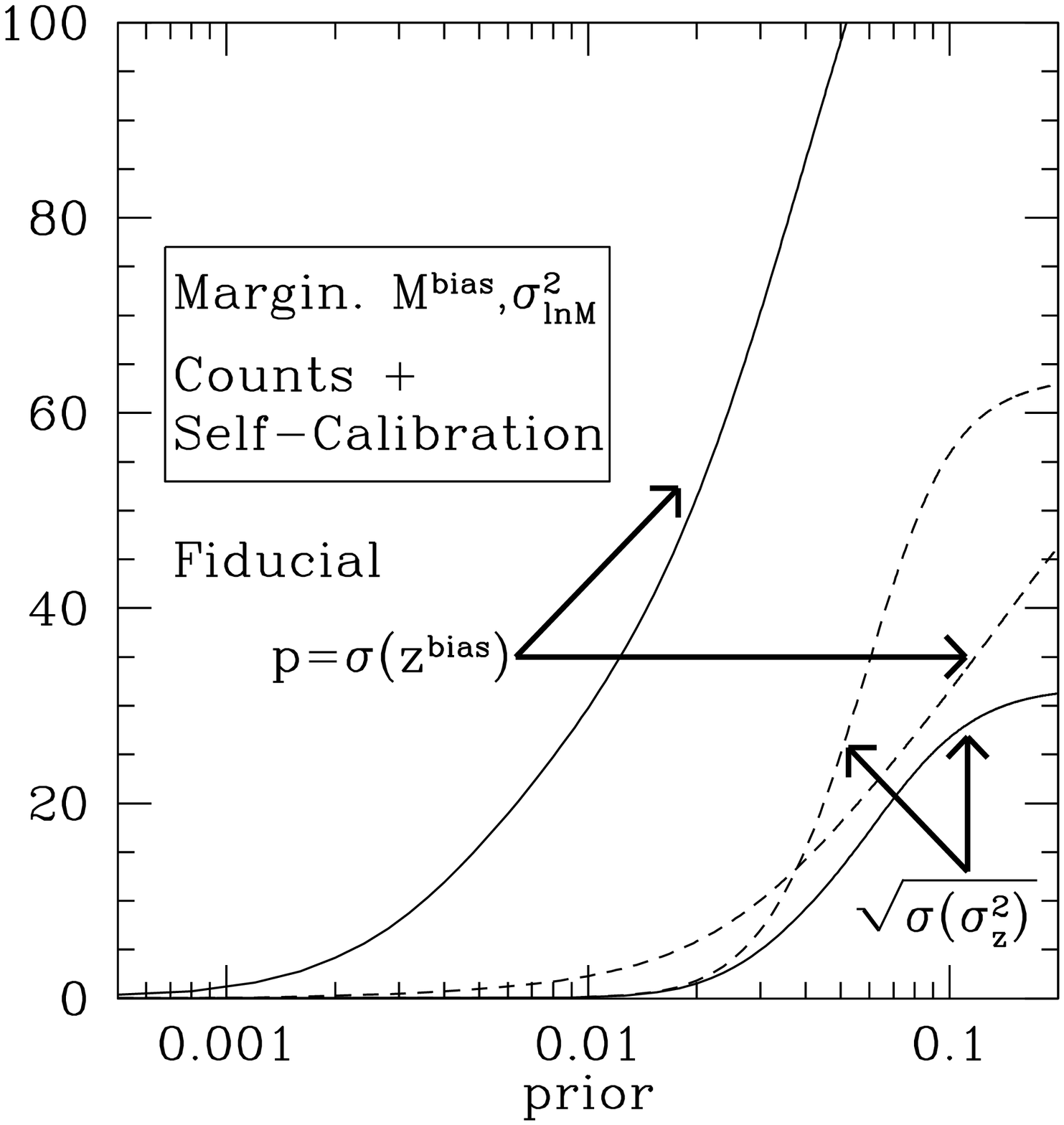}}
  \end{minipage}
  \hfill
  \begin{minipage}[t]{3.4in}
    \centerline{\epsfxsize=3.4in\epsffile{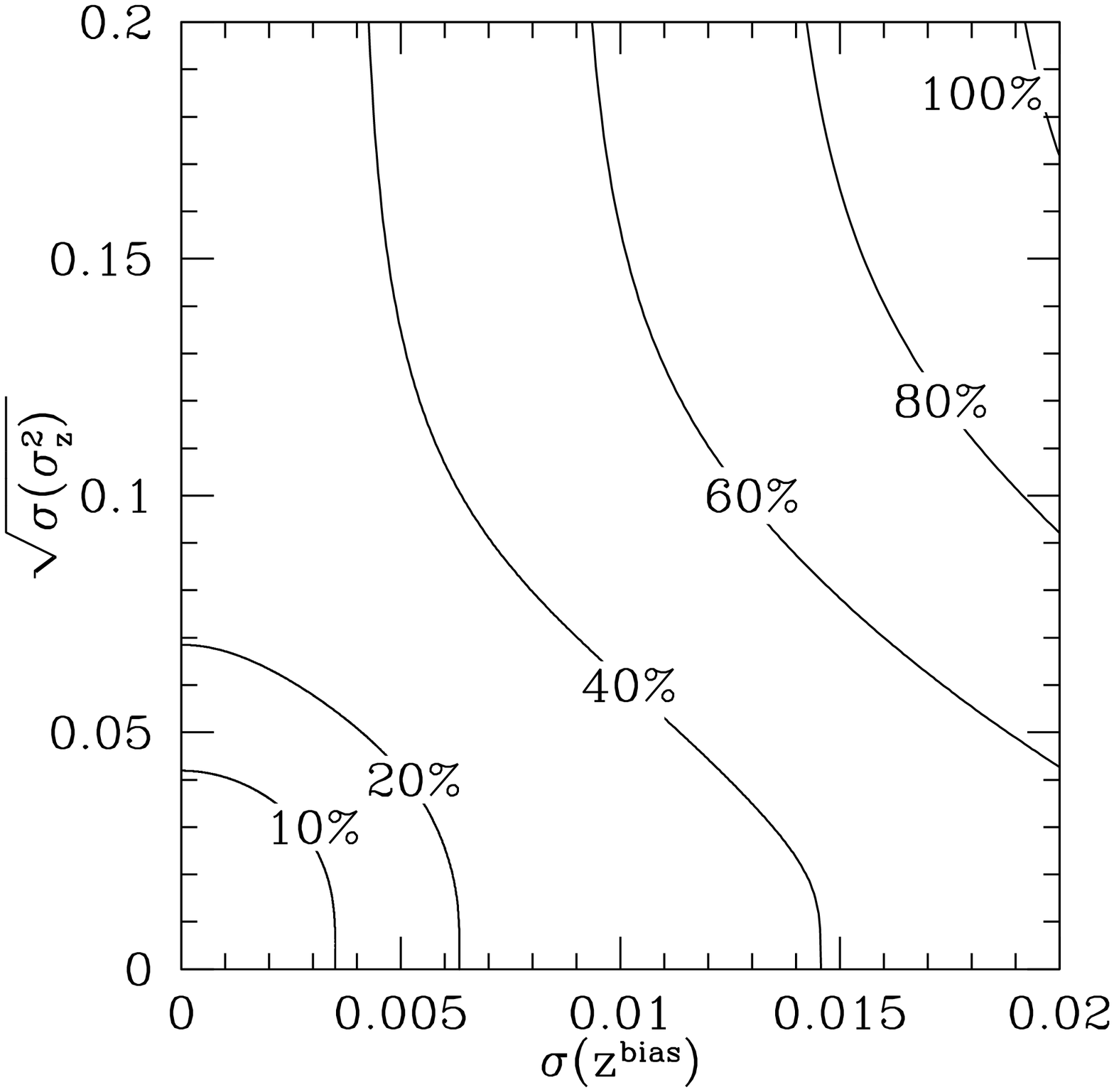}}
  \end{minipage}
\caption{\footnotesize
Dark energy sensitivity to prior knowledge of redshift parameters in the WMAP1 
cosmology and fiducial model described in \S \ref{sec_fis}. 
The {\it top row} displays results for the case of perfect masses employing
only counts information.
The {\it bottom row} is for the case of marginalized observable-mass parameters
and self-calibration with clustering and shape information. 
The {\it left panels} show percent degradation in dark energy parameter 
constraints with respect to case of perfect redshift knowledge as a function of
prior uncertainty in $\zbias$ for fixed $\sigz^2$ and as a function of
prior uncertainty in $\sigz^2$ for fixed $\zbias$.
The {\it right panels} display the respective contours of fixed degradation 
$d_w$ in the $\sigma(\zbias)$ versus  $\sqrt{\sigma(\sigz^2)}$ plane.
Compared to perfect masses, in the case of self-calibration dark energy 
degradations are weaker due to (i) worse baseline constraints and 
(ii) decreased degeneracies between dark energy and photo-z parameters, 
partially broken by self-calibration. }
\label{fig:fid.mod}
\end{figure*}

\begin{table*}[t]
  \begin{center}  
    \begin{tabular}
       {@{\extracolsep{\fill}}c|c|l|l|l|r|r}
       \hline\hline
       \multicolumn{2}{c|}{Observable-Mass parameters} & 
       \multicolumn{1}{c|}{Information} &  
       \multicolumn{2}{c|}{Constraints} &
       \multicolumn{2}{c}{Degradations}\\
       \cline{1-2}
       \cline{4-7}
       \multicolumn{1}{c|}{$\ln \Mbias$} &
       \multicolumn{1}{c|}{$\siglnM^2$} &  
       \multicolumn{1}{c|}{} &
       \multicolumn{1}{c|}{$\sigma(w)$} &
       \multicolumn{1}{c|}{$\sigma(\Omega_{\rm DE})$} &
       \multicolumn{1}{c|}{$d_{w}$} &
       \multicolumn{1}{c} {$d_{\Omega_{\rm DE}}$}\\
       \cline{1-7} 
       Known         & Known        & Counts                   & 0.033  & 0.0081 & 21\% & 17\%\\
       Known         & Marginalized & Counts                   & 0.19   & 0.21   & 90\% & 108\%\\
Marginalized         & Marginalized & Counts                   & \multicolumn{1}{c|}{-} & \multicolumn{1}{c|}{-} & \multicolumn{1}{c|}{-}& \multicolumn{1}{c}{-}\\
       \hline
       Known         & Known        & Counts + Self-Calibration& 0.031  & 0.0078 & 16\% & 15\%\\
       Known         & Marginalized & Counts + Self-Calibration& 0.072  & 0.019  &  5\% & 3\%\\
       Marginalized  & Marginalized & Counts + Self-Calibration& 0.12   & 0.029  & 11\% & 8.3\%\\
      \hline
    \end{tabular}
  \end{center}
\caption{\footnotesize 
Dark energy baseline constraints and degradations for the 
fiducial model with observable-mass parameters known/marginalized and
with/without self-calibration. 
Baseline constraints $\sigma(w)$ and $\sigma(\Omega_{\rm DE})$ 
are for $\sigma(\zbias)=\sigma(\sigz^2)=0$ and degradations $d_w$ and 
$d_{\Omega_{\rm DE}}$ relative to the baseline are for $\sigma(\zbias)=0.003$ 
and $\sqrt{\sigma(\sigz^2)}=0.03$.}
\label{tab:fid.mod}
\end{table*}

In the case of perfect knowledge of the observable-mass relation, 
the dark energy constraints from cluster counts alone are quite strong, 
as long as we have a reasonable knowledge of redshifts parameters. 
In the case of perfect redshifts, the baseline constraints are
$\sigma(w,\Omega_{\rm DE})=(0.033,0.0081)$ in the fiducial model.

These constraints are degraded as the uncertainties in the redshift
bias and variance increase. 
With counts information only, interesting dark energy constraints can 
only be extracted if we have some prior knowledge of both  redshift 
bias and scatter uncertainties, i.e. no constraints can be extracted 
if either $\sigma(\zbias)\rightarrow \infty$ or 
$\sigma(\sigz^2)\rightarrow \infty$. 

The top panels of Fig.~\ref{fig:fid.mod} show the degradations 
in dark energy for finite values of redshift uncertainties.
The top left panel shows
$d_w$ and $d_{\Omega_{\rm DE}}$ as a function of
$\sigma(\zbias)$ (with $\sigma(\sigz^2)=0$) and as a function of
$\sigma(\sigz^2)$ (with $\sigma(\zbias)=0$).
The top right panel shows contours of fixed 
$d_w$ values in the plane defined by values of 
$\sigma(\zbias)$ and $\sqrt{\sigma(\sigz^2)}$.
Table~\ref{tab:fid.mod} shows constraints and degradations in 
the fiducial model for the case of perfect masses and for the
self-calibration cases of 
\S \ref{subsec_mas.sel.cal} and \S \ref{subsec_red.sel.cal}.
As expected, the dark energy constraints are more sensitive to the 
photo-z bias compared to the photo-z scatter by about an order of 
magnitude. 

\subsection{Mass Self-Calibration} \label{subsec_mas.sel.cal}

Now we consider the case where we do not know the 
observable-mass parameters and solve for them along
with dark energy parameters simultaneously by means
of self-calibration with clustering and shape information.

\begin{figure}[tb]
   \begin{minipage}[t]{3.4in}
    \centerline{\epsfxsize=3.4in\epsffile{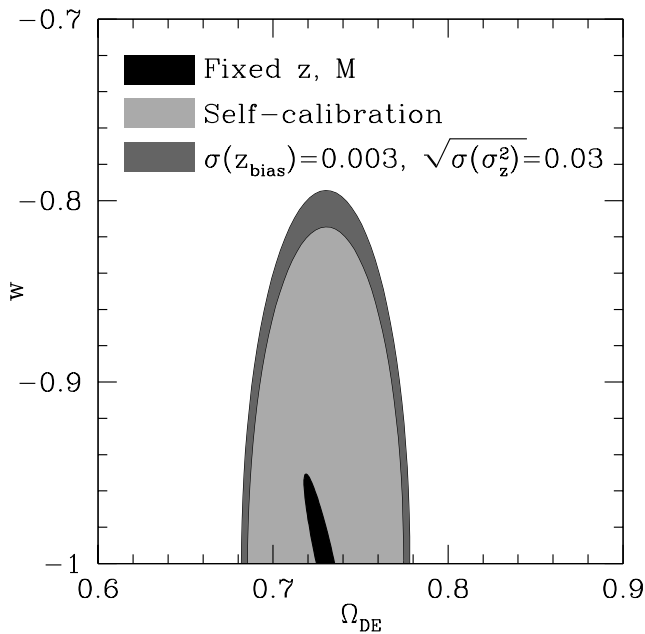}}
  \end{minipage}
  \hfill
\caption{\footnotesize
Dark energy constraints for the fiducial model in the WMAP1 cosmology. 
From inner to outer ellipses, the $68\%$ CL regions are shown for the 
case where redshift and mass parameters are perfectly known (black);
the case with self-calibration from sample covariance 
and shape information but fixed photo-z parameters (light gray); 
the case with self-calibration and prior uncertainties on photo-z parameters 
of $\sigma(\zbias)=0.003$ and $\sqrt{\sigma(\sigz^2)}=0.03$ (dark gray).
The latter case corresponds to a degradation of 
$d(w,\Omega_{\rm DE})=(11\%,8.3\%)$ with respect to the case of 
fixed redshifts.} 
\label{fig:ell.sel.cal}
\end{figure}

Let us consider first the case of perfect redshifts.
Self-calibration results are shown in Fig.~\ref{fig:ell.sel.cal} 
for our fiducial model.
For reference, the case of perfect masses is displayed in the black solid 
ellipsis corresponding to the tight baseline constraints of 
\S \ref{subsec_per.mas}. 
If no self-calibration of the observable-mass relation is used in a 
joint fit of mass and dark energy parameters, no interesting constraints 
can be extracted. 
Keeping redshifts fixed, assuming the functional forms of 
Eqs.~(\ref{eqn:biaspowerlaw}),~(\ref{eqn:variancetaylor}) for the 
mass bias and variance and employing self-calibration from clustering 
information in the sample covariance of counts produces constraints 
$\sigma(w,\Omega_{\rm DE})=(0.23,0.22)$. 
Further adding shape information from multiple mass bins allows a good 
degree of self-calibration as shown in the light gray ellipsis
of Fig.~\ref{fig:ell.sel.cal};
the constraints in this case are $\sigma(w,\Omega_{\rm DE})=(0.12,0.029)$,
i.e. there is a degradation of $d(w,\Omega_{\rm DE})=(280\%,260\%)$ with 
respect to the case of perfect mass in \S \ref{subsec_per.mas}.

As the uncertainty in redshift parameters increases, the ability to 
self-calibrate masses is reduced. 
The dark gray ellipsis in Fig.~\ref{fig:ell.sel.cal} shows how the 
self-calibrated case above is degraded if we take the photo-z parameters to have
uncertainties of $\sigma(\zbias)=0.003$ and 
$\sqrt{\sigma(\sigz^2)}=0.03$ simultaneously as opposed to being fixed; 
in this case the constraints are 
$\sigma(w,\Omega_{\rm DE})=(0.14,0.032)$, i.e. a degradation of 
$d(w,\Omega_{\rm DE})=(11\%,8.3\%)$. 
The dark energy degradation in the case of self-calibration is shown 
in the bottom panels of Fig.~\ref{fig:fid.mod} and in the bottom rows of
Table~\ref{tab:fid.mod}.
Notice that because the baseline constraints (perfect redshifts) 
are worse with self-calibration compared to fixed masses, the degradation
relative to the baseline is correspondingly weaker.
The weaker degradation is partially because the baseline constraints are
worse, but also because self-calibration helps break not only the
degeneracy between dark energy and mass, but also redshifts.
In \S \ref{subsec_red.sel.cal} we investigate that further.  

\subsection{Redshift Self-Calibration} \label{subsec_red.sel.cal}

If external mass calibrations can partially or totally constrain 
the mass bias or scatter or both, self-calibration is still useful to
monitor redshifts, therefore decreasing the redshift requirements. 
More generally, self-calibration provides important consistency checks 
that theoretical assumptions and observations agree, even more so
when we have stringent external mass priors.
External mass calibrations can come for instance from follow-up of a small 
sample of clusters with very well measured masses, weak lensing
calibrations, simulations, etc. 

Let us consider the limiting case where external calibrations put strong
priors on the mass bias, but no prior on the mass scatter, i.e.
$\sigma(\ln \Mbias)=0$ and $\sigma(\siglnM^2)=\infty$.  
In this case, without self-calibration, the constraints for perfect
redshifts are $\sigma(w,\Omega_{\rm DE})=(0.19,0.21)$ and
they degrade to $\sigma(w,\Omega_{\rm DE})=(0.35,0.44)$, i.e.
a degradation of $d(w,\Omega_{\rm DE})=(90\%,108\%)$ if we
impose redshift priors of $\sigma(\zbias)=0.003$ and 
$\sqrt{\sigma(\sigz^2)}=0.03$. 
Employing self-calibration the constraints for perfect redshifts
$\sigma(w,\Omega_{\rm DE})=(0.072,0.019)$ degrade to 
$\sigma(w,\Omega_{\rm DE})=(0.076,0.019)$ with the same
redshift priors, i.e. a degradation of only 
$d(w,\Omega_{\rm DE})=(5\%,3\%)$.

Consider now the limiting case where external priors can further
constrain the mass scatter to $\sigma(\siglnM^2)=0$ in addition
to the mass bias.
Without self-calibration, the constraints for perfect
redshifts are $\sigma(w,\Omega_{\rm DE})=(0.033,0.0081)$ and they 
degrade to $\sigma(w,\Omega_{\rm DE})=(0.039,0.095)$ if we
impose redshift priors of $\sigma(\zbias)=0.003$ and 
$\sqrt{\sigma(\sigz^2)}=0.03$. , i.e. a degradation of 
$d(w,\Omega_{\rm DE})=(21\%,17\%)$. 
Employing self-calibration the constraints for perfect redshifts
$\sigma(w,\Omega_{\rm DE})=(0.031,0.0078)$ degrade to 
$\sigma(w,\Omega_{\rm DE})=(0.036,0.0090)$ with the same
redshift priors, i.e. a degradation of 
$d(w,\Omega_{\rm DE})=(16\%,15\%)$.
These results are also displayed in Table~\ref{tab:fid.mod}.
Interestingly, in the limit where we have no redshift knowledge
($\sigma(\zbias)=\sigma(\sigz^2)=\infty$), without self-calibration
no interesting dark energy constraints are possible, 
but employing full self-calibration we can still
obtain $\sigma(w,\Omega_{\rm DE})=(0.11,0.11)$.

These results show that self-calibration is very effective not
only in allowing a joint fit of mass and cosmological parameters, but
also in reducing the redshift requirements in those cases.
As external priors put constraints on some of the observable-mass relation
parameters, but not all of them, self-calibration becomes even more important.
In the limit where external priors are even stronger and masses are perfectly
determined, self-calibration becomes less important, but it still helps
monitor redshifts.
However, even in this case prior redshift knowledge is still
important in order to extract the survey full constraining power.
Since any given survey is likely to be in between these limit cases,
both self-calibration techniques and good knowledge of redshifts
will be comparably important.

\subsection{Fiducial Model Dependence} \label{subsec_fid}

In this section we consider deviations from our fiducial model
and their effect on the dark energy degradation results.
Even though changes in the fiducial values of the nuisance
parameters describing the observable-mass relation and photo-z's have
a very small effect on the constraints, changes in their
parametrization will obviously affect the results. 
For instance, increasing the number of parameters in the mass-observable 
relation to allow for more general models produces worse baseline 
constraints and correspondingly weaker degradations due to photo-z 
uncertainties.
The constraints obviously also change if the fiducial cosmology and survey
specifications deviate considerably from the ones assumed in the
fiducial model. 
We will first consider a fiducial cosmology based on cosmological parameters 
from WMAP3 and also changes in our fiducial survey properties.
Finally we will consider deviations in the photo-z model assumed.

\subsubsection{Fiducial Cosmology and Survey Properties} \label{subsubsec_cos}

\begin{figure}[tb]
  \begin{minipage}[t]{3.4in}
    \centerline{\epsfxsize=3.4in\epsffile{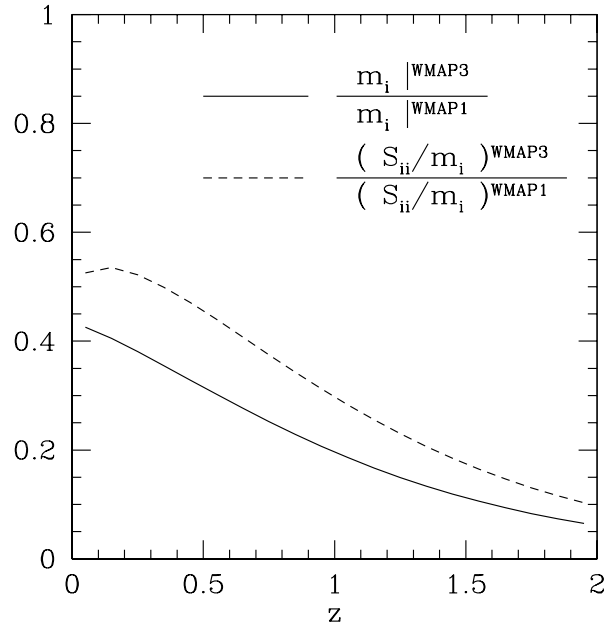}}
  \end{minipage}
\caption{\footnotesize
The lower value of $\sigma_8$ in the WMAP3 cosmology decreases
the number counts with respect to WMAP1, specially at high redshift. 
Similarly, the relative importance of sample variance errors with 
respect to the increased shot noise decreases, lowering the power 
of using sample variance as a signal for self-calibration.
} 
\label{fig:numvarW1W3}
\end{figure}

Because of the exponential sensitivity of the cluster mass function to the 
growth of structure, the decrease in the value of $\sigma_8$ for WMAP3 causes 
the overall number counts to be reduced by a factor of $\sim 4$ with respect 
to the WMAP1 case. 
Likewise, the reduction in counts decreases the relative importance of 
sample variance errors with respect to Poisson shot noise. 
In Fig.~\ref{fig:numvarW1W3} we show the ratio of the average number counts 
$m_i(z)$ in the WMAP3 and WMAP1 cosmologies, as well as the ratio of the 
corresponding 
sample variance $S_{ii}$ relative to Poisson variance $m_i$.

Let us first consider the case of perfect knowledge of the observable-mass 
parameters. 
In the case of perfect redshifts, the baseline constraints for the WMAP3 
cosmology with count information become 
$\sigma(w,\Omega_{\rm DE})=(0.047,0.010)$. 
Even though the WMAP3 total number
counts are reduced by a factor of $\sim 4$ compared to WMAP1,
the baseline constraints degrade only by 
$d(w,\Omega_{\rm DE})=(44\%,25\%)$. 
If all clusters carried the same cosmological information, we would 
naively expect a factor of $\sim 2$ degradation from the increased Poisson 
errors.
However, the rare high mass/redshift clusters are more sensitive
to cosmology. The fact that the relative reduction on the number of
these rare clusters is more pronounced provides extra cosmology information 
that partially balances 
the degradation from the overall counts reduction.

\begin{figure}[tb]
\begin{minipage}[t]{3.4in}
    \centerline{\epsfxsize=3.4in\epsffile{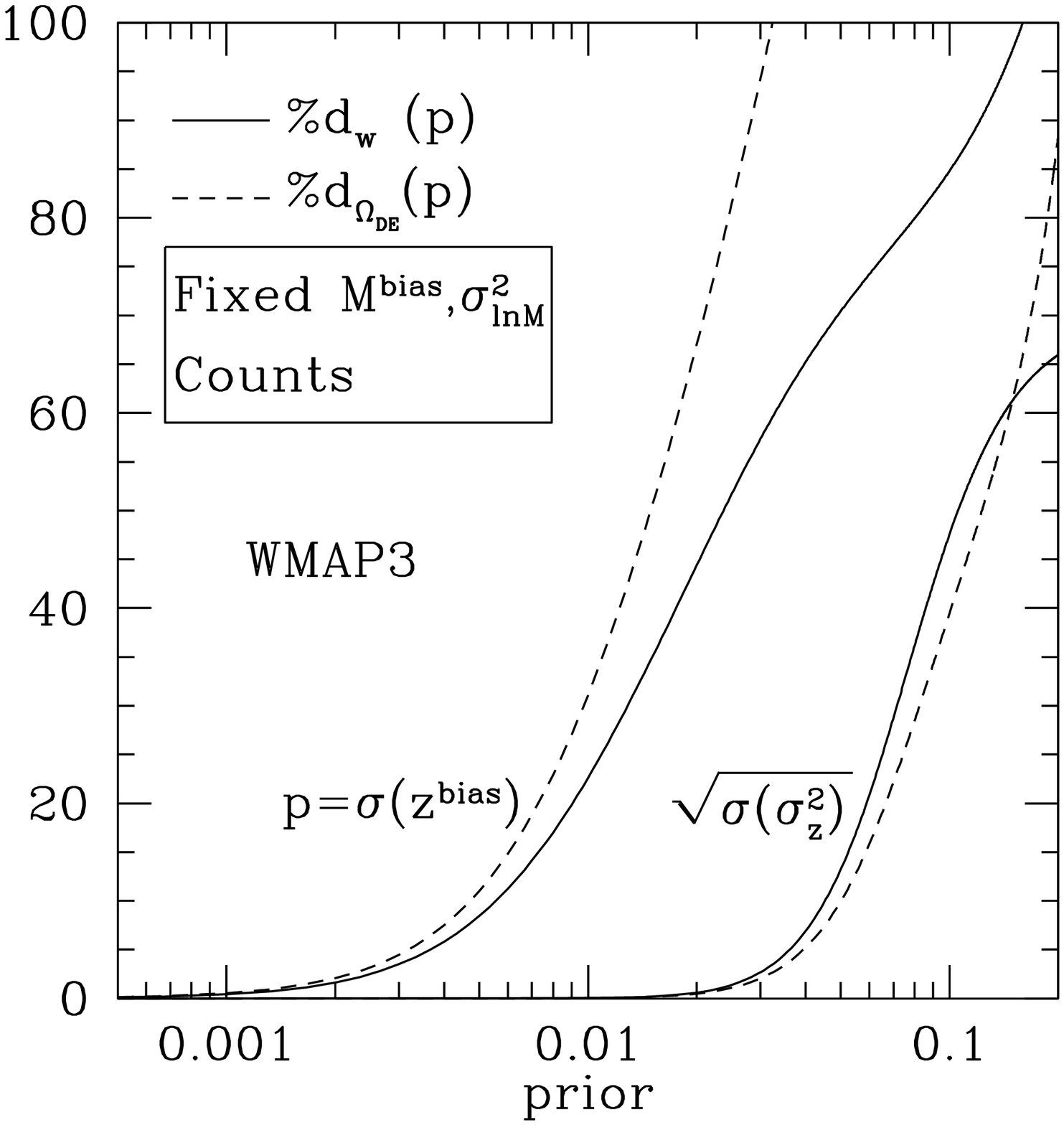}}
  \end{minipage}
  \begin{minipage}[t]{3.4in}
    \centerline{\epsfxsize=3.4in\epsffile{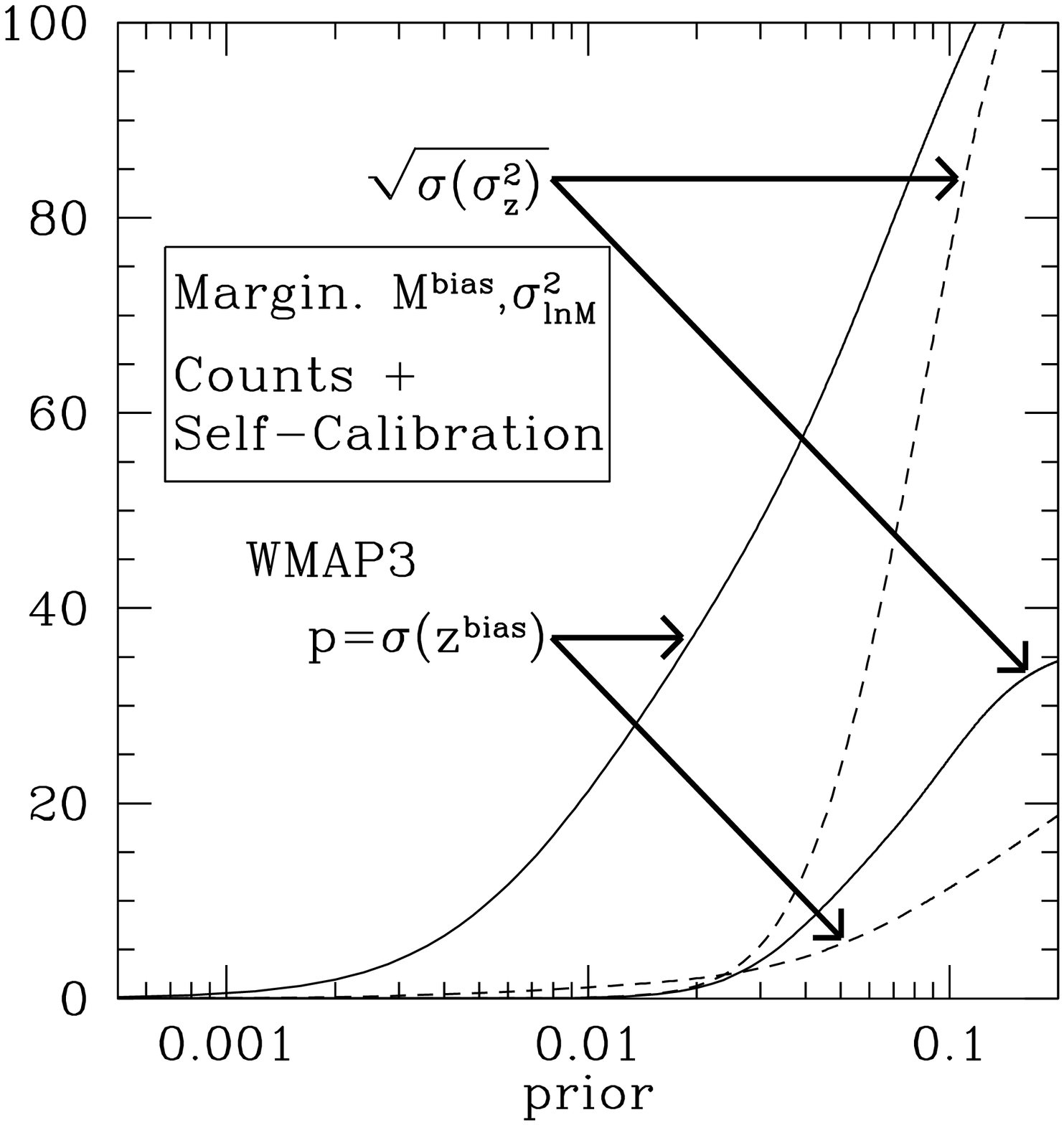}}
  \end{minipage}
  \hfill
\caption{\footnotesize
Degradation in dark energy parameters similar to Fig.~\ref{fig:fid.mod} but 
for the WMAP3 cosmology. 
}
\label{fig:W3}
\end{figure}

\begin{table*}[t]
  \begin{center}  
    \begin{tabular}
       {@{\extracolsep{\fill}}c|c|c|l|l|l|r|r}
       \hline\hline
       \multicolumn{1}{c|}{Case} &
       \multicolumn{1}{c|}{Number Counts} &
       \multicolumn{1}{c|}{Observable-Mass} & 
       \multicolumn{1}{c|}{Information} &  
       \multicolumn{2}{c|}{Constraints} &
       \multicolumn{2}{c}{Degradations}\\
       \cline{5-8}
       \multicolumn{1}{c|}{} &  
       \multicolumn{1}{c|}{per deg$^2$} &
       \multicolumn{1}{c|}{parameters} &
       \multicolumn{1}{c|}{} &
       \multicolumn{1}{c|}{$\sigma(w)$} &
       \multicolumn{1}{c|}{$\sigma(\Omega_{\rm DE})$} &
       \multicolumn{1}{c|}{$d_{w}$} &
       \multicolumn{1}{c} {$d_{\Omega_{\rm DE}}$}\\
       \cline{1-8} 
        Fiducial model                      & 8.0 & Known         & Counts                    & 0.033  & 0.0081 & 21\%   & 17\%\\
        WMAP3                               & 2.1 & Known         & Counts                    & 0.047  & 0.010  &  5.7\% &  6.1\%\\
$\Mobs_{\rm th}=10^{14.435}h^{-1}M_{\odot}$ & 2.1 & Known         & Counts                    & 0.050  & 0.012  &  5.6\% &  5.8\%\\
        Area = 1000 deg$^2$                 & 2.0 & Known         & Counts                    & 0.054  & 0.014  &  4.9\% &  3.5\%\\
        $\zphot_{\rm max}=0.5$              & 2.3 & Known         & Counts                    & 0.063  & 0.015  &410\%   &780\%\\
         \hline
        Fiducial model                      & 8.0 & Marginalized  & Counts + Self-Calibration & 0.12  & 0.029 & 11\%   & 8.3\%\\
        WMAP3                               & 2.1 & Marginalized  & Counts + Self-Calibration & 0.24  & 0.048 &  7.0\% & 6.2\%\\
$\Mobs_{\rm th}=10^{14.435}h^{-1}M_{\odot}$ & 2.1 & Marginalized  & Counts + Self-Calibration & 0.24  & 0.052 &  7.3\% & 5.4\%\\
        Area = 1000 deg$^2$                 & 2.0 & Marginalized  & Counts + Self-Calibration & 0.24  & 0.052 &  4.0\% & 2.9\%\\
        $\zphot_{\rm max}=0.5$              & 2.3 & Marginalized  & Counts + Self-Calibration & 0.28  & 0.034 & 35\%   & 42\%\\
      \hline
    \end{tabular}
  \end{center}
\caption{\footnotesize 
Dark energy baseline constraints and degradations for the fiducial model
(WMAP1, $\Mobs_{\rm th}=10^{14.2}h^{-1}M_{\odot}$, Area = 4000 deg$^2$,
$\zphot_{\rm max}=2.0$) 
and  various deviations
from the fiducial survey properties and cosmology. 
In all cases the total number of clusters is 
reduced by a factor of $\sim 4$ with respect to the fiducial case.
Baseline constraints are for $\sigma(\zbias)=\sigma(\sigz^2)=0$ 
and relative degradations are for $\sigma(\zbias)=0.003$ and
$\sqrt{\sigma(\sigz^2)}=0.03$.}
\label{tab:dev.cosm.surv}
\end{table*}

Let us now consider the dark energy degradations due to photo-z 
uncertainties. 
The top panel of Fig.~\ref{fig:W3} shows the degradation of dark 
energy parameters as a function of photo-z uncertainties for the WMAP3 
cosmology. 
Notice that dark energy degradations relative to the baseline cases  
are slightly weaker for the WMAP3 case, since its baseline is worse, 
confirming the trend that surveys with smaller yields require 
less knowledge of redshift parameters for fixed baseline degradations. 
In other words, in the limit of perfect masses for the WMAP3 case, 
dark energy constraints are less sensitive to photo-z uncertainties 
because the best achievable results are worse.

Now we consider the case where we marginalize over the observable-mass
parameters and employ self-calibration. 
In this case, with perfect redshifts the constraints for 
WMAP3 are $\sigma(w,\Omega_{\rm DE})=(0.24,0.048)$, i.e. a degradation of
$d(w,\Omega_{\rm DE})=(410\%,370\%)$ with respect to the case of perfect
masses and $d(w,\Omega_{\rm DE})=(96\%,64\%)$ with respect to the 
corresponding self-calibrated case in the WMAP1 cosmology.
Self-calibration is based on the ability to divide the
cluster sample in space and mass. 
The decrease in the number counts for WMAP3 reduces the relative 
importance of sample variance over the increased shot noise, and 
reduces the power of self-calibration by clustering. 
In addition, the ability to split the sample in mass bins is 
decreased because we run out of clusters at high masses, and 
self-calibration from shape information is also reduced.  

The bottom panel of Fig.~\ref{fig:W3} shows the dark energy 
degradation as a function of photo-z uncertainties for the case of 
self-calibration in the WMAP3 cosmology.
Relative to the case of perfect masses and to the self-calibrated case
of WMAP1, most degradations for WMAP3 are weaker, as expected from the 
worse baseline constraints.
However the $\sigma(\Omega_{\rm DE})$ degradation as
a function $\sqrt{\sigma(\sigz^2)}$ is stronger relative to both cases.
That happens because self-calibration uses
sample covariance information, which is exponentially sensitive
to $\sigz^2$. 
With the reduced relative importance of sample
variance in WMAP3, changes in $\sigz^2$ are now relatively more
degenerate with cosmology and observable-mass parameters. 

To make the point clearer, let us consider the case where we
employ self-calibration only from shape information, but not
clustering. 
In this case the self-calibration constraints
for perfect redshifts are
$\sigma(w,\Omega_{\rm DE})=(0.26,0.11)$, i.e. the $w$ constraint
is nearly identical to the full self-calibration case, but
the $\Omega_{\rm DE}$ constraint is still $\sim 2$ times higher.
Keeping the photo-z bias fixed
$\sigma(\zbias)=0$ and applying a prior on the photo-z variance of
$\sqrt{\sigma(\sigz^2)}=0.2$, the
constraints degrade to $\sigma(w,\Omega_{\rm DE})=(0.87,0.17)$, 
i.e. a degradation of $d(w,\Omega_{\rm DE})=(240\%,54\%)$.
With full self-calibration the corresponding degradation is
$(30\%,110\%)$. 
Therefore whereas clustering self-calibration from sample variance 
does not help the $w$ constraint
to further decrease considerably from its value with shape 
self-calibration only, it does make this constraint less sensitive 
to photo-z uncertainties. 
On the other hand, clustering self-calibration improves the 
$\Omega_{\rm DE}$ constraint, but
leaves it more unstable to photo-z scatter uncertainties.
These features tell us that, when employing self-calibration,
knowledge of the photo-z scatter is essential, specially in
a situation of fewer clusters such as in the WMAP3 cosmology.

For completeness, we consider other survey modifications that lead to
roughly the same reduction in the total number counts as the WMAP3 cosmology
relative to WMAP1. 
Table~\ref{tab:dev.cosm.surv} shows results for the fiducial model along
with the WMAP3 case and changes in the mass threshold $\Mobs_{\rm th}$,
total survey area, and maximum survey redshift $\zphot_{\rm max}$. 
All cases lead to a reduction in the total counts by a factor of $\sim 4$ 
relative to the fiducial model. In the case where we change the 
maximum redshift to $\zphot_{\rm max}=0.5$, we also reduce the number
of photo-z parameters to $N_z=5$ so as to keep one value of $\zbias$
and $\sigz^2$ per photo-z bin of $\delta z^p_i=0.1$.
Except for the change in $\zphot_{\rm max}$, which makes dark energy 
constraints and degradations much worse, all other cases produce results 
roughly similar.
Obviously, although these changes produce similar counts reduction, they all
remove different clusters. 
For instance, the WMAP3 reduction is more pronounced at high redshifts, 
whereas the change in $\Mobs_{\rm th}$ removes all low mass clusters and 
the change in survey area reduces the numbers of all clusters equally.

In the case of perfect masses and counts only, except for the $z_{\rm max}$
reduction case, the cases with worse baseline constraints have weaker
degradations; 
the more drastic change of reducing $\zphot_{\rm max}$ has a much higher 
impact because with the total removal of high redshift clusters, 
the leverage to probe the evolution of the growth of structure is 
severely diminished.

In the case of self-calibration, many factors affect the degradations.
For the fiducial model and the cases of decrease in area and $\zphot_{\rm max}$, 
the degradations are smaller compared to the case of only counts, because
their baseline constraints are much worse and self-calibration is fairly
efficient. 
Even though the degradation in the case of $\zphot_{\rm max}$ reduction is 
still the highest, it is many times smaller than in the case of counts only, 
because with this small redshift range, self-calibration is able to break 
most of the high degeneracy between mass, redshift and cosmology.
For the cases of WMAP3 and reduction in $\Mobs_{\rm th}$, the degradations
are larger compared to counts only and perfect masses. In the WMAP3 case
self-calibration is harder because of the reduction in $\sigma_8$,
which lowers the value of $S_{ii}$. In the case of reduction in 
$\Mobs_{\rm th}$, the removal of the low mass clusters makes 
self-calibration from shape information harder.
In any case, we confirm the trend that more ambitious surveys will be able
to extract more information from self-calibration techniques but will also
require more redshift knowledge for a full realization of their
constraining power.

\subsubsection{Fiducial Observable-Mass Parameters} \label{subsubsec_obs}

\begin{figure}[tb]
 \begin{minipage}[t]{3.4in}
   \centerline{\epsfxsize=3.4in\epsffile{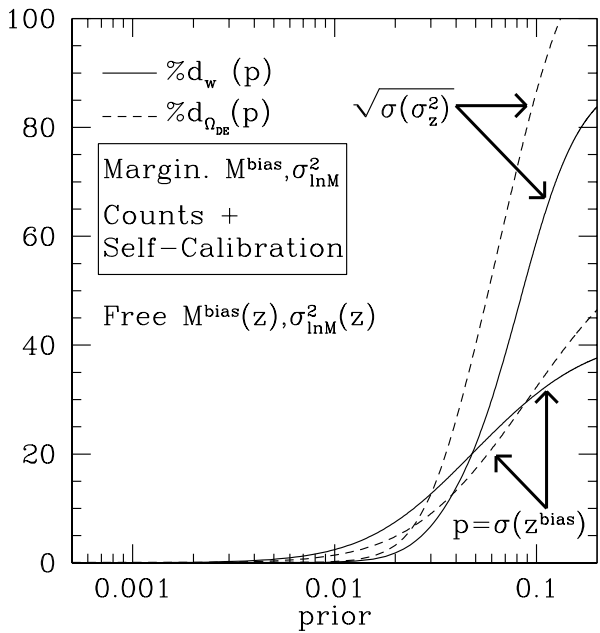}}
 \end{minipage}
\caption{\footnotesize
Degradation in dark energy parameters for the case where $\Mbias$ and 
$\siglnM^2$ are free functions
of redshift as opposed to having the functional forms of
Eqs.~(\ref{eqn:biaspowerlaw}),~(\ref{eqn:variancetaylor}) 
assumed in our fiducial model.}
\label{fig:dev.mass.obs}
\end{figure}

Now we return to the WMAP1 cosmology case, but we change our fiducial
parametrization of the observable-mass relation. 
We allow the mass bias and variance to be general functions of redshift, 
i.e. instead of assuming the functional forms in
Eqs.~(\ref{eqn:biaspowerlaw}) 
and (\ref{eqn:variancetaylor}), we now have one value 
of $\Mbias$ and $\siglnM^2$ in each photo-z bin of $\delta z^p_i=0.1$. 
For the case of fixed mass parameters, the dark energy constraints 
obviously do not change, but they become 
$\sigma(w,\Omega_{\rm DE})=(0.22,0.030)$, i.e. a degradation of
$d(w,\Omega_{\rm DE})=(80\%,4\%)$ with respect to the 
 fiducial model with self-calibration.

In Fig.~\ref{fig:dev.mass.obs} we show the corresponding dark energy 
degradation from the baseline constraints as a function of photo-z 
uncertainties in this case. 
Now $\sigma(w)$ has a worse baseline, and its degradation is less 
sensitive to $\sigma(\zbias)$ but more sensitive to $\sigma(\sigz^2)$
compared to the fiducial model.
On the other hand, since the baseline $\sigma(\Omega_{\rm DE})$ 
basically did not change, meaning that it is making strong use of 
self-calibration, its degradation as a function of $\sigma(\zbias)$
is nearly identical, but is much stronger as a function of $\sigma(\sigz^2)$.  
In this conservative case of $40$ parameters for the
observable-mass relation, where self-calibration is being
used in its full power, if we are interested in moderate 
dark energy degradations ($\sim 10\%-20\%$), both 
$\sigma(\zbias)$ and $\sigma(\sigz^2)$ are
important. As we transit to the case with fewer parameters,
$\sigma(\zbias)$ becomes relatively more important.

\subsubsection{Fiducial Photo-z Parameters} \label{subsubsec_pho}

\begin{figure}[tb]
 \begin{minipage}[t]{3.4in}
    \centerline{\epsfxsize=3.4in\epsffile{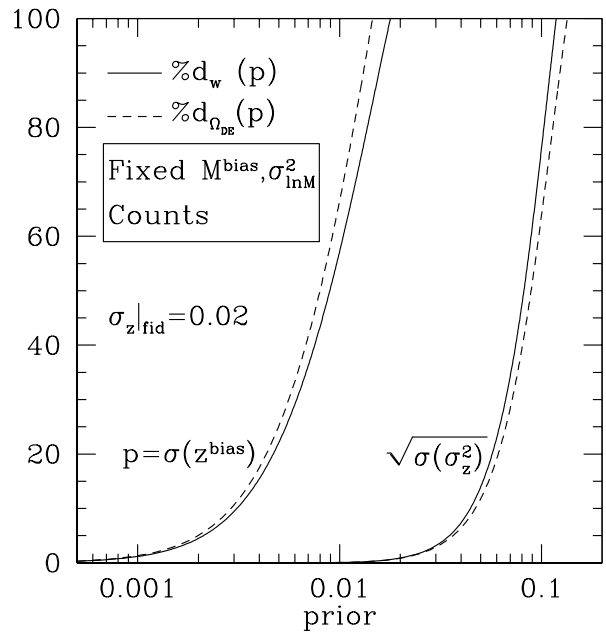}}
  \end{minipage}
  \begin{minipage}[t]{3.4in}
    \centerline{\epsfxsize=3.4in\epsffile{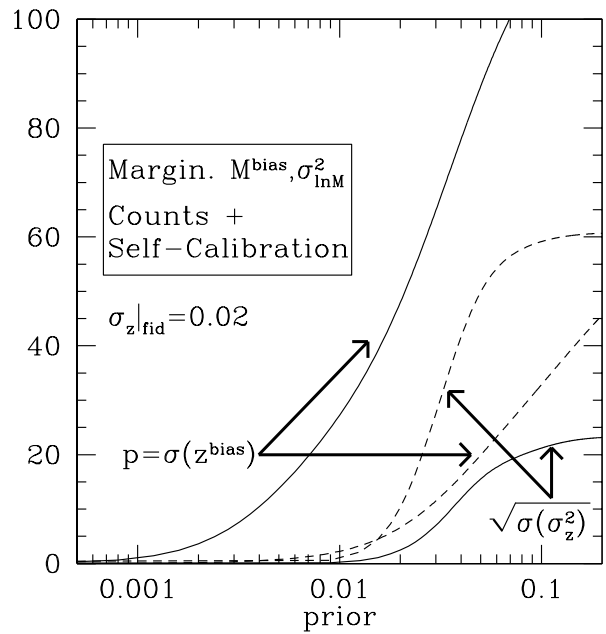}}
  \end{minipage}
\caption{\footnotesize
Dark energy parameters degradation for a fiducial photo-z scatter of 
$\sigz |_{\rm fid}=0.02$ constant in redshift as opposed to the 
fiducial model of Eq.~(\ref{eqn:sigzlinear}).
}
\label{fig:dev.sigz}
\end{figure}

Finally, we consider the case where Eq.~(\ref{eqn:sigzlinear}) in our 
fiducial model is replaced by a fiducial photo-z scatter of 
$\sigz |_{\rm fid}=0.02$ constant in redshift.
Let us consider the case of perfect masses first. In the case of perfect 
redshifts the baseline constraints in this case are 
$\sigma(w,\Omega_{\rm DE})=(0.032,0.080)$, i.e. they remain nearly the same 
as in our fiducial case ($\sim 1\%$ improvement).
In the top panel of Fig.~\ref{fig:dev.sigz} we show dark energy 
degradations as we allow for uncertainties in photo-z parameters. 
Since the $\zbias$ model is still the same, 
degradations as a function of $\sigma(\zbias)$ are
identical to the fiducial case. As a function of $\sqrt{\sigma(\sigz^2)}$
they are only slightly weaker.

In the case of self-calibration, the baseline constraints are
$\sigma(w,\Omega_{\rm DE})=(0.11,0.026)$, i.e. they improved by
$(6\%,10\%)$ with respect to the corresponding self-calibrated case
in the fiducial model. Now the improvement is more noticeable
than in the case of perfect masses since we use  
sample variance to self-calibrate mass parameters.
In the bottom panel of Fig.~\ref{fig:dev.sigz} we show dark energy 
degradations in this case.
Again, degradations as functions
of $\sigma(\zbias)$ are identical to the fiducial model. 
As a function of $\sqrt{\sigma(\sigz^2)}$ degradations become stronger
at lower values of $\sigma(\sigz^2)$, since we are now in a better
baseline. At higher values of $\sigma(\sigz^2)$, degradations become
weaker, since self-calibration is now more efficient.

\begin{table*}[t]
  \begin{center}  
    \begin{tabular}
       {@{\extracolsep{\fill}}c|c|l|l|l|r|r}
       \hline\hline
       \multicolumn{1}{c|}{Case} &
       \multicolumn{1}{c|}{Observable-Mass} & 
       \multicolumn{1}{c|}{Information} &  
       \multicolumn{2}{c|}{Constraints} &
       \multicolumn{2}{c}{Degradations}\\
       \cline{4-7}
       \multicolumn{1}{c|}{} &  
       \multicolumn{1}{c|}{parameters} &
       \multicolumn{1}{c|}{} &
       \multicolumn{1}{c|}{$\sigma(w)$} &
       \multicolumn{1}{c|}{$\sigma(\Omega_{\rm DE})$} &
       \multicolumn{1}{c|}{$d_{w}$} &
       \multicolumn{1}{c} {$d_{\Omega_{\rm DE}}$}\\
       \cline{1-7}
        Fiducial model                & Known         & Counts                    & 0.033  & 0.0081 & 21\% & 17\%\\
        $\sigz |_{\rm fid}=0.02$      & Known         & Counts                    & 0.032  & 0.0080 & 12\% & 13\%\\
        $N_z=10$ and $\delta z^p=0.1$ & Known         & Counts                    & 0.033  & 0.0081 & 15\% & 15\%\\
        $N_z=20$ and $\delta z^p=0.2$ & Known         & Counts                    & 0.033  & 0.0083 & 51\% & 52\%\\
        $N_z=10$ and $\delta z^p=0.2$ & Known         & Counts                    & 0.033  & 0.0083 & 14\% & 14\%\\
         \hline
        Fiducial model               & Marginalized  & Counts + Self-Calibration & 0.12  & 0.029 & 11\% &  8.3\%\\
        $\sigz |_{\rm fid}=0.02$     & Marginalized  & Counts + Self-Calibration & 0.11  & 0.026 & 15\% & 30\%\\
        $N_z=10$ and $\delta z^p=0.1$& Marginalized  & Counts + Self-Calibration & 0.12  & 0.029 & 13\% & 14\%\\
        $N_z=20$ and $\delta z^p=0.2$& Marginalized  & Counts + Self-Calibration & 0.13  & 0.031 & 22\% &  6.1\%\\
        $N_z=10$ and $\delta z^p=0.2$& Marginalized  & Counts + Self-Calibration & 0.13  & 0.031 & 11\% &  4.8\%\\
      \hline
    \end{tabular}
  \end{center}
\caption{\footnotesize 
Dark energy baseline constraints and degradations for the 
fiducial model ($\sigz |_{\rm fid}=0.03(1+z)$, $N_z=20$ and $\delta z^p=0.1$)
and deviations from the fiducial photo-z parameters. 
Baseline constraints are for $\sigma(\zbias)=\sigma(\sigz^2)=0$ 
and relative degradations are for $\sigma(\zbias)=0.003\sqrt{N_z/20}$ and
$\sigma(\sigz^2)=(0.03)^2\sqrt{N_z/20}$. 
The priors are scaled to reflect constraints per fixed bin 
$\delta z =0.1$ of {\it true} redshift.}
\label{tab:dev.sigz}
\end{table*}

Finally, let us consider the effect of changing the number
of photo-z parameters $N_z$ and/or the photo-z bin size
$\delta z^p$. 
In Table~\ref{tab:dev.sigz} we show the fiducial model
and variations in the photo-z model that include
changes in the fiducial values of the photo-z scatter,
$N_z$ and $\delta z^p$. 
The degradations are shown for priors of
$\sigma(\zbias)=0.003\sqrt{N_z/20}$ and
$\sigma(\sigz^2)=(0.03)^2\sqrt{N_z/20}$.
The scaling of the priors with $N_z$ comes from 
Eqs.~(\ref{eqn:calib_zbias}),(\ref{eqn:calib_sigz})
and the fact that the number $N_{\rm spec}$ of spectroscopic 
calibrators in a given redshift region is inversely proportional to the 
number of redshift bins $N_{\rm spec}\propto 1/N_z$.
These scaled priors then reflect constraints per 
fixed bin $\delta z =0.1$ of {\it true} redshift independently 
of the value of $N_z$ used. 
Notice from Table~\ref{tab:dev.sigz} that the baseline constraints
do not change much by doing this changes in the fiducial model
but the degradations relative to the baseline are affected
by a series of competing effects.

Decreasing the number of photo-z parameters to $N_z=10$ while
keeping the photo-z bin size at $\delta z^p_i=0.1$ (i) partially 
decreases some degeneracies between dark energy and the fewer
photo-z parameters and (ii) makes the photo-z parameters
smoother functions of redshift, making photo-z
perturbations more nonlocal (less restricted to a particular photo-z bin).
The latter effect causes (a) the counts sensitivity to $\zbias$ to increase
because it depends on the value of $\zbias$ at the bin borders, and 
(b) the sensitivity to $\sigz^2$ decreases because the number
of objects scattered up, controlled by the values of $\sigz$ at the
bin borders, more efficiently compensates the number of 
objects scattered down, controlled by $\sigz$ at the bin center. 
In the case of counts only and perfect mass the dark energy degradations
are weaker than the fiducial case whereas in the case of unknown masses
and self-calibrations they are stronger.

Increasing the bin size to $\delta z^p=0.2$ but keeping $N_z=20$
(i) produces a coarser probe of cluster mass function but (ii)
reduces the Poisson noise in the counts, therefore increasing
the importance of sample variance and clustering self-calibration.
Notice though that in all cases absolute errors increase with $\delta z^p$ 
since subdividing the data can only help constraints.

If in addition to using coarser bins of $\delta z^p=0.2$
we simultaneously decrease the number of photo-z parameters to
$N_z=10$ so as to still have one photo-z parameter per photo-z bin, 
the degradations in all cases become slightly better than in the 
fiducial model, although around slightly worse baselines. 
That shows that our results are relatively robust to the particular 
binning choice if we keep the number of photo-z parameters per 
bin of true redshift constant.  
In this case, coarser photo-z bins lose shape information
on the redshift distribution but allows the fewer photo-z parameters 
to be better known by the same factor that they were reduced in number.
Since the best  achievable constraints are worse with such wider
bins and the power of cluster counts surveys lies exactly in
their ability to split the counts in a reasonable number of 
redshift and mass bins, photo-z bins much coarser than these 
are unlikely to be adopted in future cluster survey analyses 
if they are to be competitive dark energy probes.

\subsection{Training Set Requirements}

The photo-z requirements for a fixed dark energy degradation
ultimately translate into requirements on the size of the calibration 
training set (with known redshifts), from which we infer the photo-z 
error parameters.
With the assumption made here that the cluster photo-z errors have a
Gaussian distribution, the uncertainties in the photo-z
parameters $\zbias(z_i)$ and $\sigz^2(z_i)$ in a given redshift 
bin $i$ are related to the number $N_{\rm spec}(z_i)$ of spectroscopic 
training set calibrators in that bin by
\begin{eqnarray}
\sigma(\zbias(z_i))&=&\sigz(z_i)\sqrt{\frac{1}{N_{\rm spec}(z_i)}}\,,
\label{eqn:calib_zbias} \\
\sigma(\sigz^2(z_i))&=&\sigz^2(z_i)\sqrt{\frac{2}{N_{\rm spec}(z_i)}}\,.
\label{eqn:calib_sigz}
\end{eqnarray}
   
We are interested in the values of $N_{\rm spec}$ necessary in order
not to degrade dark energy constraints appreciably. 
There are basically two ways to go about this question, which depend
on the choice of the spectroscopic calibrators. 
The first case corresponds to choosing the calibrators to be clusters 
with at least one member galaxy whose spectroscopic redshift 
has been measured, in which case $N_{\rm spec}$
corresponds to the number of such spectroscopic clusters and
$\sigz(z_i)$ is the cluster photo-z scatter in the redshift bin.
This choice is natural, but may suffer from the fact that
a small number of such clusters are likely to be available in a given survey.
The second choice then tries to compensate for that by including more 
objects in the calibration process. 
Since most clusters contain large numbers of red elliptical
galaxies and optical cluster finder methods are typically tuned to 
find exactly such clusters, one can try to use not only the cluster
spectroscopic redshifts, but the redshifts of all spectroscopic
elliptical galaxies including field galaxies.
Although this would make use of a much larger sample of objects,
they would typically have a larger photo-z scatter and the assumption
that all red galaxies (in the field and in clusters) share the same
photo-z error properties would have to be checked.
In any case, the first choice is general enough in the sense that we
can think of the red galaxy sample as being used, not directly
in the cluster photo-z parameters calibration, but in the 
estimation of these parameters. 
From that perspective, having a large spectroscopic 
sample of red galaxies helps calibrate the photo-z's of cluster
member galaxies and decrease the cluster photo-z scatter, 
since each cluster member provides roughly an independent photo-z 
measurement of the cluster photo-z. 
Even though some cluster finding methods, such as those based on the
SZ signal and weak lensing shear, do not preferentially detect clusters
with large numbers of red galaxies, most future cluster surveys will
have a spectroscopic follow-up of a sample of the detected clusters.
For that reason, in the following analysis we will 
take the first choice of redshift calibrators, 
i.e. spectroscopic clusters.
   
Recall in Fig.~\ref{fig:fid.mod} we considered fixed degradations in the 
dark energy equation of state $w$, given priors  $\sigma(\zbias)$
and $\sigma(\sigz^2)$ which were  {\it constant} in redshift. 
In that case, each value of $d_w$ requires
$\sigma(\zbias)$ and $\sigma(\sigz^2)$ to be lower than certain 
values, which in turn produce lower limits on $N_{\rm spec}(z_i)$. 
For illustrative purposes let us fix $d_w=10\%$ and consider the 
fiducial model with counts information only, in which case 
Fig.~\ref{fig:fid.mod} requires 
$\sigma(\zbias)<0.003$ and $\sqrt{\sigma(\sigz^2)}<0.03$.
The condition on $\sigma(\zbias)$, which is more stringent than the one on
$\sigma(\sigz^2)$, requires the number of calibrators in bins of 
$\delta z_i=0.1$ to be 
$N_{\rm spec}(z_i=0.5) > 110$ and $N_{\rm spec}(z_i=1.95) > 870$.  
However, all these numbers are excessively high
due to our requirement of constant $\sigma(\zbias)$ 
and $\sigma(\sigz)$ in each bin, including the high redshift ones that
have little impact on the dark energy constraints. 

To remedy this problem, let us instead take $N_{\rm spec}$ to be a fixed
fraction of the total number in a given bin
\begin{eqnarray}
N_{\rm spec}(z_i) = am_i \,.
\label{eqn:def.a}
\end{eqnarray}
We can then study the requirements on dark energy degradations
as a function of $a$.
We should keep in mind though that, observationally, it is even harder 
to obtain spectra for the few clusters at high redshifts 
compared to those (also few ones) at very low redshift, because
at higher redshifts the $4000${\rm \AA} break of elliptical galaxies leaves
the optical filter coverage. Therefore in  
practice $a$ will probably be a decreasing function of redshift.

\begin{figure}[tb]
 \begin{minipage}[t]{3.4in}
    \centerline{\epsfxsize=3.4in\epsffile{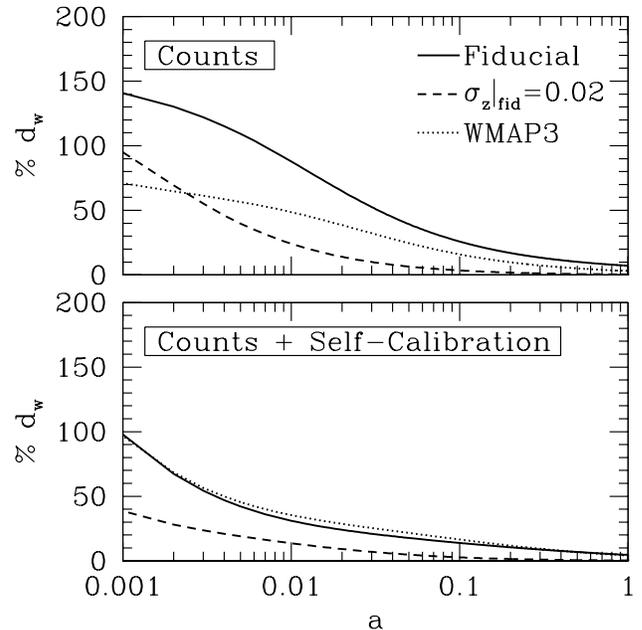}}
  \end{minipage}
\caption{\footnotesize
Dark energy parameters degradation in the dark energy equation of
state as a function of the fraction $a$ of clusters with spectroscopic
redshifts for the fiducial model, WMAP3 cosmology and photo-z scatter of 
$\sigz |_{\rm fid}=0.02$, using counts information only and
self-calibration.
}
\label{fig:deg.a}
\end{figure}
 
In Fig.~\ref{fig:deg.a}, we show the degradation $d_w$ as a 
function of $a$ in the fiducial model and also for the WMAP3 
cosmology as well as the case with $\sigz|_{\rm fid}=0.02$.
Notice that even in the limit where all clusters have photo-z's
($a=1$), because the number of these clusters is finite, we still 
have some degradation compared to the case where the photo-z 
parameters have zero uncertainties. 
Even though this {\it intrinsic} degradation can be as high as
$\sim 10\%$, as shown in the top panel of Fig.~\ref{fig:deg.a}, if
$a$ was actually equal to 1, one would not use photo-z's in the
first place, since all objects would have perfect redshift 
measurements. Therefore actual degradations are smaller than those
shown in Fig.~\ref{fig:deg.a} for very high values of $a$.
Also, in practice only values of $a$ that realize integer numbers of 
clusters are allowed. However, since we are considering $a$ constant
in redshift, we used continuous values of $a$, even those that give
$N_{\rm spec} < 1$. Thus, actual degradations 
for very low values of $a$ are larger than those shown in 
Fig.~\ref{fig:deg.a}, especially for WMAP3.

Ignoring these caveats just mentioned, we see from the top panel in 
Fig.~\ref{fig:deg.a} that the WMAP3 case has a smaller degradation 
than the fiducial model when employing counts only, consistently with 
the results of \S \ref{subsubsec_cos}. 
Moreover reducing the photo-z scatter has an important impact on 
the $w$ degradation as expected.

From the bottom panel of Fig.~\ref{fig:deg.a}, we see that 
self-calibration tends to make the training size requirements 
smaller in most cases. 
Interestingly, $d_w$ in both the fiducial model
and the WMAP3 cosmology are nearly identical as a function of $a$.
That happens because for fixed values of $\sigma(\zbias(z_i))$ and 
$\sigma(\sigz^2(z_i))$, whereas $w$ degradations are typically
weaker for WMAP3, the numbers $m_i$ of existing clusters  
are smaller and consequently the number of spectroscopic calibrators
are reduced by the same factor, making it harder to achieve 
these photo-z accuracies.
 
These results highlight the importance of having the best
photo-z estimators available, which allow for the photo-z scatter
to be as small as possible; this has a direct impact in reducing the
required calibration set size. 
In the context of self-calibration, having spectroscopic redshifts 
for only $\sim 1\%$ of the clusters is enough to keep 
$d_w \lesssim 30\%$, whereas in the case of
counts only, this spectroscopic sample needs to contain at least
$\sim 8\%$ of the clusters.
For a more stringent requirement of $d_w \lesssim 20\%$, with 
self-calibration it is necessary to have spectroscopic
redshifts for $\sim 4\%$ of the clusters. 
Quantitatively, this requires $\sim 1300$ spectroscopic clusters in 
the fiducial model and $\sim 300$ in the WMAP3 cosmology. 
Given that the training set size requirements for other cosmological
probes such as weak lensing and baryon acoustic oscillations are on
the order of hundreds of thousands, only a few percent of these 
galaxies need to be in clusters in order for those training sets to
serve also as cluster calibration samples, without additional
observational effort. 
Complementing the existing training sets with spectroscopic
follow-up observations would then further improve dark energy constraints
and allow for degradations at the percent level.

\subsection{Caveats}

There are a number of caveats associated with our results.  We have
taken a constant $\Mobs_{\rm th}$ whereas in reality we expect it
 to be cosmology and 
redshift dependent for a SZ survey like SPT (although less than in an 
X-ray survey \cite{HaiMohHol01}). 
Our constant fiducial $\Mobs_{\rm th}$ can be thought of as an 
effective mass threshold that produces the same number counts as the 
true threshold.
If the effective mass threshold in a real survey deviates considerably 
from the one considered here, the baseline dark energy constraints will 
change, and so will the degradations with respect to this baseline.
As discussed in \S \ref{sec_res}, a change in $\Mobs_{\rm th}$ has an 
effect similar to a change in $\sigma_8$ or in the survey area, 
namely the change in the total cluster counts.

The variation of the observable-mass relation with cosmology 
 may also open up 
parameter degeneracies.    
For our fiducial model we took the observable-mass distribution to be
parametrized by $\ln \Mbias$ and $\siglnM^2$ with a fixed functional form 
in redshift.    If this form instead mimicked the effect of a cosmological
parameter, a degeneracy would appear.
A complete treatment of these issues, which is beyond the scope of this
work, would require the choice of a specific observable and extensive 
simulations to obtain the true parameter values and their distributions.

The potential impact of degeneracies can be
assessed by considering 
the case 
 where $\ln \Mbias$ and $\siglnM^2$ are allowed to 
be arbitrary functions of redshift (\S \ref{subsec_fid}). 
In this case the baseline 
dark energy constraints degrade by up to $80\%$ relative
to the fiducial model.  
Degradations on top of that due to uncertainties in 
$\zbias$ and $\sigz^2$ become weaker and stronger 
respectively. 

If one further includes an arbitrary mass dependence of these parameters, 
the techniques explored here are not sufficient to provide a good degree of 
self-calibration, even with perfect redshifts. 
Our results would still be valid as long as trends in mass are 
well-known close to the mass threshold \cite{LimHu05}. 
Recent simulations \cite{KraVikNag06,Nag06,Ohaetal06} suggest that 
observable-mass relations can actually be parametrized with scaling relations
involving few parameters as assumed here. 
Since our main purpose here is to illustrate the 
 {\it relative} degradation of 
dark energy constraints due to photo-z errors, the range of models considered 
here should suffice. 

Finally, we assumed the photo-z error parameters to follow a Gaussian 
distribution and depend on redshift but not mass.
This assumption does not hold for field galaxies binned 
only in redshift due to the mixing of various galaxy types.
Even though we expect Gaussian errors to be a good approximation 
for clusters, which have mostly red elliptical galaxies and lower 
photo-z scatter, possible non-Gaussian features induced by
membership contamination may require proper modeling.

The photo-z parameters, in particular the photo-z scatter 
is expected to have some mass dependence.
More massive clusters have more galaxies to average over when 
computing the cluster photo-z, allowing the photo-z scatter
to decrease.
We envision that the redshift/mass dependence of photo-z parameters
can be modeled from mock catalogs and cast in simple scaling 
relations, likely depending on the particular cluster finder/survey.
In the lack of any specific modeling, we chose to keep the 
redshift dependence general and ignore the mass dependence
of the photo-z parameters. 
Lastly, we did not take into account possible errors in the theoretical 
cluster-mass function, which might introduce parameter biases or 
interact differently with photo-z errors.

\section{Discussion} \label{sec_dis}

The use of cluster abundance as a competitive cosmological
probe requires strict knowledge of both the cluster 
observable-mass relations and photo-z's. 
Self-calibration techniques help break degeneracies 
between cosmology and cluster masses and, at the same
time, decrease redshift knowledge requirements. 
When external mass calibrations are available, self-calibration
may become even more important, assisting monitor
redshifts and providing interesting consistency checks.
Conversely, good photo-z knowledge is required in order
to extract optimal constraints from self-calibration methods.

We studied the effect of photo-z uncertainties in dark energy
constraints assuming a fiducial model with survey specifications
similar to the SPT, a constant mass threshold $\Mobs_{\rm th}$,
a mass bias $\ln \Mbias$ parametrized by a power law in redshift,
a mass variance $\siglnM^2$ with a cubic redshift evolution and
photo-z parameters assumed to be general functions of redshift
but not mass.

For this fiducial model it will be necessary to have priors of
$\sigma(\zbias) \lesssim 0.003$ and $\sqrt{\sigma(\sigz^2)} \lesssim 0.03$ 
in the case of perfect masses and counts information in order not to 
degrade dark energy constraints by more than $\sim 10 \%$. 
In the case of self-calibration of the observable-mass relation,
these priors uncertainties should be kept at
$\sigma(\zbias) \lesssim 0.004$ and $\sqrt{\sigma(\sigz^2)} \lesssim 0.04$.
These requirements become weaker/stronger if we increase/decrease the
number of nuisance parameters describing masses and redshifts or
if we consider less/more ambitious surveys or cosmologies yielding
less/more clusters.

In order to achieve these requirements, it is important not only
to use the best general photo-z techniques, but also to understand the 
cluster finding selection functions and contamination fractions; otherwise 
even if the photo-z methods work well for field galaxies, the
cluster redshifts and masses might be misestimated.
Large and representative training sets help decrease the 
redshift uncertainties and keep the dark energy constraints close to
their baseline values and also help self-calibration be more effective.
From the self-calibration perspective, simulations play a very important
role. Self-calibration techniques are more effective when we can parametrize
the observable-mass relation with a small number of parameters, typically with
simple scaling relations. Simulations do not need to determine exact values of
these parameters, but only provide a good confidence
that these simple relations are stable to theoretical uncertainties
and have relatively small theoretical scatter. If that is the case
one can extract these parameter values from self-calibration methods, which
use information that naturally comes in surveys without 
requiring additional observational efforts. 

Photo-z methods are expected to be ever improving during the
evolution of ongoing and 
future surveys such as the Panoramic Survey Telescope and Rapid Response
System (PanSTARRS)~\cite{Kaietal02}, the SPT~\cite{Ruh04}, 
the Dark Energy Survey (DES)~\cite{Abbetal05}, 
the Supernova Acceleration Probe (SNAP)~\cite{Ald05} and
the Large Synoptic Survey Telescope (LSST)~\cite{Tys02}.
There are good prospects for the existence 
of large and representative training sets that will allow these surveys
to be reasonably well calibrated. 
Likewise we expect simulations to improve our knowledge of the dark matter 
halo mass-functions and of the various observable-mass relations.
External calibrations from weak lensing are likely to also play an important 
role in mass determinations. 
Finally, self-calibration methods can further help
reduce requirements and provide important consistency checks that
theoretical assumptions have been or not realized in the observations.
All these aspects will be important and complementary in consistent
cosmological analyses of future cluster surveys.
 
\smallskip

{
\noindent{\it Acknowledgments:}  We thank Carlos Cunha, Josh Frieman, Huan Lin, 
Zhaoming Ma, and Hiroaki Oyaizu 
for useful discussions.  
This work was supported by the DOE and the KICP
under NSF PHY-0114422.  WH was additionally supported  by the 
David and Lucile Packard Foundation.}
\smallskip

\appendix 

\section{Window Function} \label{app_win}

Recall we used spherical coordinates ${\bf x}=(r,\theta,\phi)$ to 
parametrize the position vector describing the window volume element
such that $d^3x = r^2 d\Omega dr$ with $d\Omega=\sin\theta d\theta d\phi$.
The cell defines a solid angle given by $\Delta \Omega=\pi \theta_s^2$ in 
the small angle approximation. 
For convenience, we will also use cylindrical coordinates to parametrize
the same vector ${\bf x}=(r,\rho,\phi)$, where $r(z)$ is still the angular
diameter distance to redshift $z$ and now the perpendicular coordinates
$(\rho,\phi)$ parametrize the angular extension of the window. In these
coordinates we have $d^3x = \rho d\rho d\phi dr$ and the cell defines 
a extension of $\rho_s=r\theta_s$. We split the total 
window function into $W_i({\bf x})=W^{\rm th}_{\perp i}(\rho,\phi ) F_i(z)$.
Here and below the superscript ``th" denotes a top hat function whose
value is unity within the pixel and zero outside of the pixel.

The volume-normalized Fourier transform of the window is given by
\begin{eqnarray}
W_i(\bk)&=&\frac{1}{V_i}\int d^3 x e^{i{\bf k.x}  } W_i({\bf x}) \nonumber \\
        &=& \frac{1}{V_i}W_{\perp i}^{\rm th}({\bf \kperp})W_{\parallel i}(\kpar)\,,
\label{eqn:win_split}
\end{eqnarray}
\noindent where the window volume is 
\begin{eqnarray}
V_i \equiv \int{d^3 x} W_i(\bx) 
       = \Delta \Omega \int dr r^2 F_i(z) 
\end{eqnarray}
\noindent and we defined
\begin{eqnarray}
W_{\perp i}^{\rm th}({\bf \kperp}) &\equiv& \int d\rho d\phi \rho e^{i \kperp \rho \cos\phi } W_{\perp i}^{\rm th}(\rho,\phi) \,,\\
W_{\parallel i}(\kpar) &\equiv&\int dr e^{i \kpar r} F_i(z)\,.
\label{eqn:win_par} 
\end{eqnarray}
Note that the normalized Fourier window $W_i(\bk)\rightarrow 1$ as $\bk \rightarrow 0$.

Similarly, in the case of a top hat total window
we would have
\begin{eqnarray}
W_i^{\rm th}(\bk)= \frac{1}{V_i^{\rm th}}W_{\perp i}^{\rm th}({\bf \kperp})W_{\parallel i}^{\rm th}(\kpar)\,,
\label{eqn:win_split_th}
\end{eqnarray}
\noindent with
\begin{eqnarray}
V_i^{\rm th} &\equiv& \int{d^3 x} W_i^{\rm th}(\bx) 
                 =\Delta \Omega \int dr r^2 W_{\parallel i}^{\rm th}(r) \,.
\end{eqnarray}

From Eqs.~(\ref{eqn:win_split}) and (\ref{eqn:win_split_th}) we may write
\begin{eqnarray}
\frac{ W_i(\bk) }{ W_i^{\rm th}(\bk) }=\frac{ V_i^{\rm th} }{ V_i }
                                       \frac{ W_{\parallel i}(\kpar) }{ W_{\parallel i}^{\rm th}(\kpar) }\,,
\label{eqn:win_ratio}
\end{eqnarray}
\noindent and we can rewrite Eq.~(\ref{eqn:win_par}) as
\begin{eqnarray}
W_{\parallel i}(\kpar)&=&\int \frac{dz}{H(z)} e^{i \kpar r(z)} F_i(z)\,.
\end{eqnarray}

We will assume that $H(z)$ does not change
appreciably in the photo-z bin and, because $F_i(z)$ quickly drops outside
the bin, the values of $r$ that contribute to the integral are restricted
to the range $ r=r_i \pm \delta r_i/2$ with 
$\delta r_i \approx  \delta z^p_i/H_i$ and $H_i \equiv H(z^p_i)$, so that 
\begin{eqnarray}
W_{\parallel i}(\kpar)\approx \frac{1}{H_i}\int dz e^{i \kpar z/H_i} F_i(z)
          =\frac{F_i(k_{z,i})}{H_i}\,,
\label{eqn:win_par_z}
\end{eqnarray}
\noindent where we defined $k_{z,i}=\kpar/H_i$. Note that $F_i(z)$ can be 
expressed as a convolution of the Gaussian selection with a top hat window 
in redshift
\begin{eqnarray}
F_i(z)=\int d\zphot W_{\parallel i}^{\rm th}(\zphot) p(\zphot| z)\,.
\end{eqnarray}

The convolution theorem gives 
$F_i(k_{z,i})=W_{\parallel i}^{\rm th}(k_{z,i}) p(k_{z,i})$ 
so that Eq.~(\ref{eqn:win_par_z}) becomes
\begin{eqnarray}
W_{\parallel i}(\kpar)=\frac{W_{\parallel i}^{\rm th}(k_{z,i}) p(k_{z,i})}{H_i}\,.
\end{eqnarray}
By a similar argument that led to Eq.~(\ref{eqn:win_par_z}) we have that 
$W_{\parallel i}^{\rm th}(k_{z,i}) \approx H_i W_{\parallel i}^{\rm th}(\kpar)$ 
and Eq.~(\ref{eqn:win_ratio}) becomes
\begin{eqnarray}
\frac{W_i(\bk)}{W_i^{\rm th}({\bk})} =  \frac{ V_i^{\rm th} }{ V_i } p(k_{z,i})\,.
\end{eqnarray}

In the limit where $r(z)$ and the photo-z parameters $\zbias$ and
$\sigz^2$ do not change appreciably inside the photo-z bin, the
window volumes in the absence ($V_i^{\rm th}$) and presence of photo-z
errors ($V_i$) roughly coincide 
$V_i \approx V_i^{\rm th} \propto r_i^2 \delta r_i$, i.e. the photo-z
errors distort the volume element but do not change its value.
Therefore we have
\begin{eqnarray}
W_i(\bk)= W_i^{\rm th}({\bk}) p(k_{z,i}).
\end{eqnarray}
The Fourier transform of the top hat window is \cite{HuKra02}
\begin{eqnarray}
W_i^{\rm th}({\bk})&=& 2e^{i \kpar r_i} 
              \frac{\sin( \kpar \delta r_i/2) }{\kpar \delta r_i/2} 
              \frac{J_1(\kperp r_i \theta_s)}{\kperp r_i \theta_s} \label{hukrav02}
\end{eqnarray}
\noindent and for the Gaussian distribution we have
\begin{eqnarray}
p(k_{z,i}) &=& e^{ik_{z,i} \zbias_i} e^{-\sigma_{z,i}^2 k_{z,i}^2/2}.
\end{eqnarray}
Combining these results, we obtain 
\begin{eqnarray}
W_i(\bk)&=&2 \exp{\left[ i \kpar \left( r_i+\frac{\zbias_i}{H_i} \right) \right] }
              \exp{ \left(- \frac{\sigma_{z,i}^2 \kpar^2}{2H_i^2} \right) } \nonumber \\
           && \frac{\sin( \kpar \delta r_i/2) }{\kpar \delta r_i/2} 
              \frac{J_1(\kperp r_i \theta_s)}{\kperp r_i \theta_s}\, 
\nonumber 
\label{window}
\end{eqnarray}
for the window function.
\vfill

\bibliography{clusterphotoz}

\end{document}